\documentclass[a4paper,12pt]{article}
\usepackage{jheppub}
\usepackage[dvipsnames]{xcolor}
\usepackage{tikz}
\usepackage{pgfplots}
\usepackage[T1]{fontenc}
\usepackage[]{slashed}
\usepackage[]{bm}     
\usepackage{physics}
\usepackage{dsfont}
\usepackage[compat=1.1.0]{tikz-feynman}
\usetikzlibrary{shapes.geometric}
\usepackage[force]{feynmp-auto}

\usepackage{hyperref}
\hypersetup{colorlinks=true,linkcolor=magenta,anchorcolor=green,citecolor=cyan,filecolor=black,menucolor=black,urlcolor=brown}
\makeatletter

\usepackage{lipsum}

\usepackage{graphicx}
\usepackage{epstopdf}
\usepackage{bbm,amsmath,graphicx,amssymb,amsfonts,amsthm}

\usepackage{soul}
\definecolor{bubbles}{rgb}{0.91, 1.0, 1.0}
\definecolor{aquamarine}{rgb}{0.5, 1.0, 0.83}
\definecolor{bubblegum}{rgb}{0.99, 0.76, 0.8}
\definecolor{blackbell}{rgb}{0.64, 0.64, 0.82}
\definecolor{dollarbill}{rgb}{0.72, 0.93, 0.6}

\def\sq[#1,#2]{\left[#1\,#2\right]}
\def\an[#1,#2]{\left\langle#1\,#2\right\rangle}

\def\an[#1,#2]{\left\langle#1\,#2\right\rangle}
\def\aq[#1,#2,#3]{\left\langle#1|#2|#3\right]}
\def\qa[#1,#2,#3]{\left[#1|#2|#3\right\rangle}
\def\sq[#1,#2]{\left[#1\,#2\right]}
\def\spa#1.#2{\left\langle#1\,#2\right\rangle}
\def\spab[#1,#2,#3]{\left\langle#1|#2|#3\right]}
\def\spba[#1,#2,#3]{\left[#1|#2|#3\right\rangle}
\def\spb#1.#2{\left[#1\,#2\right]}

\def\Ttrma(#1,#2,#3,#4){{\rm tr}_{-}[\slash \!\!\!\;\!\! #1\slash  \!\!\!\;\!\! #2 \slash  \!\!\!\;\!\!#3\slash  \!\!\!\;\!\!#4]}
\def\Ttrmb(#1,#2,#3,#4,#5,#6){{\rm tr}_{-}[\slash \!\!\!\;\!\! #1\slash  \!\!\!\;\!\! #2 \slash  \!\!\!\;\!\!#3\slash  \!\!\!\;\!\!#4\slash  \!\!\!\;\!\!#5\slash  \!\!\!\;\!\!#6]}
\def\Ttrmc(#1,#2,#3,#4,#5,#6,#7,#8){{\rm tr}_{-}[\slash \!\!\!\;\!\! #1\slash  \!\!\!\;\!\! #2 \slash  \!\!\!\;\!\!#3\slash  \!\!\!\;\!\!#4\slash  
\!\!\!\;\!\!#5\slash  \!\!\!\;\!\!#6\slash  \!\!\!\;\!\!#7\slash  \!\!\!\;\!\!#8]}
\def\Dp(#1,#2){(#1\cdot #2)}

\def\triangleboxleft{\scalebox{.9}{$\triangleleft$}\kern-.1em\Box}
\def\triangleboxright{\Box\kern-.1em\scalebox{.9}{$\triangleright$}}
\def\dBox{\Box\kern-.1em\Box}
\def\dNPBoxs{\scalebox{.9}{$\bowtie$}\kern-.1em\Box}
\def\dNPBoxu{\Box\kern-.1em\scalebox{.9}{$\bowtie$}}

\def\beq{\begin{equation}}
\def\eeq{\end{equation}}
\def\bes{\begin{split}}
\def\ees{\end{split}}
\def\beqa{\begin{eqnarray}}
\def\eeqa{\end{eqnarray}}

\def\eeqa{\end{eqnarray}}
\def\ek[#1,#2]{(\varepsilon_{#1}\cdot k_{#2})}
\def\e[#1,#2]{(\varepsilon_{#1}\cdot \varepsilon_{#2})}
\def\s(#1,#2){{(\ell_#1\cdot\ell_#2)}}

\def\e{\epsilon}

\pgfdeclarelayer{bg}    
\pgfsetlayers{bg,main}  

\definecolor{Mathematica}{HTML}{ed192d}

\tikzset{graviton/.style={decorate, decoration={snake, amplitude=.4mm, segment length=1.5mm, pre length=.5mm, post length=.5mm}, double}}

\usepackage{float}

\usetikzlibrary{shapes.misc}
\usetikzlibrary{arrows}
\usetikzlibrary{decorations.markings}
\usetikzlibrary{positioning,fit}
\usetikzlibrary{patterns}

\tikzset{cross/.style={cross out, draw=black, minimum size=2*(#1-\pgflinewidth), inner sep=0pt, outer sep=0pt},
	cross/.default={2pt}}

 \preprint{CERN-TH-2023-096, IPhT-t23/040, LAPTH-023/23}
      
\title{The Relation Between KMOC and Worldline Formalisms for Classical Gravity}

\author[a,b]{\!\! Poul H. Damgaard}
\author[a]{\!\!, Elias Roos Hansen}
\author[a]{\!\!, Ludovic Plant\'e}
\author[c]{\!\!, Pierre Vanhove}

\affiliation[a]{Niels Bohr International Academy, Niels Bohr Institute, University of Copenhagen, Blegdamsvej 17, DK-2100 Copenhagen, Denmark}
\affiliation[b]{Theoretical Physics Department, CERN, 1211 Geneva 23, Switzerland}
\affiliation[c]{Institut de Physique Theorique, Universit\'e Paris-Saclay,
CEA, CNRS, F-91191 Gif-sur-Yvette Cedex, France}

\abstract{We demonstrate the equivalence between observables in the KMOC and worldline formalisms for classical general relativity, highlighting the relation between the initial conditions in the two frameworks and how the Keldysh-Schwinger in-in formalism is contained in both of them even though the KMOC representation conventionally leads to the evaluation of
scattering amplitudes with Feynman propagators. The relationship between the two approaches is illustrated in detail for the momentum kick at second Post-Minkowskian order.}

\begin{document} 
\maketitle
\flushbottom
\newpage
\section{Introduction}\label{sec:intro}

The amplitude-based approach to the Post-Minkowskian expansion of general relativity has proven to be remarkably efficient,
and in a rapid sequence of steps it has led to a complete solution up to third Post-Minkowskian order~\cite{Damour:2016gwp,Damour:2017zjx,Bjerrum-Bohr:2018xdl,Cheung:2018wkq,Bern:2019nnu,Bern:2019crd,Damour:2019lcq,DiVecchia:2020ymx,Damour:2020tta,DiVecchia:2021ndb,DiVecchia:2021bdo,Bjerrum-Bohr:2021vuf,Bjerrum-Bohr:2021din,Bjerrum-Bohr:2022blt,Manohar:2022dea}. At fourth
Post-Minkowskian order~\cite{Bern:2021dqo,Bern:2021yeh,Bjerrum-Bohr:2022ows,DiVecchia:2022owy,DiVecchia:2022piu} 
the amplitude method has so far provided most contributions, although not all terms associated with gravitational back-reaction from
radiation have been computed yet using that framework. In the probe limit (where radiative effects can be ignored), results have already been presented up to fifth Post-Minkowskian order~\cite{Bjerrum-Bohr:2021wwt}.

A number of new amplitude approaches have been suggested as the calculations have entered new and unexplored territory. This includes
an effective Hamiltonian prescription for the conservative parts~\cite{Cheung:2018wkq,Cristofoli:2019neg,Cristofoli:2020uzm,Kalin:2019rwq,Bjerrum-Bohr:2019kec} and, 
most notably, the KMOC-formalism for the computation of classical observables from quantum field theory 
\cite{Kosower:2018adc,Maybee:2019jus,Herrmann:2021lqe,Herrmann:2021tct,Cristofoli:2021vyo,Cristofoli:2021jas}. Inspired by the fourth-order subtraction scheme explored in ref.~\cite{Bern:2021dqo} an
amplitude formalism based on an exponential representation of the scattering matrix $S$ was suggested in ref.~\cite{Damgaard:2021ipf}. It is covariant and has the advantage that it by construction starts with the classical contribution to the phase, followed by
quantum corrections. It thus seems perfectly suited for a semiclassical expansion and the appearance of superclassical terms 
due to a rewriting in terms of the conventional $T$-matrix as in 
\beq
\hat{S} = 1 + \frac{i}{\hbar}\hat{T} \label{StoT}
\eeq
are guaranteed to cancel. One can thus exclusively work with those
parts that contain the classical pieces, ignoring all other terms. An alternative formulation based on the large-mass expansion
field theory amplitudes~\cite{Damgaard:2019lfh,Aoude:2020onz} has also
been
advocated~\cite{Brandhuber:2021eyq,Brandhuber:2021bsf,Brandhuber:2022enp,Brandhuber:2023hhy}
alternatively, 
by performing a multi-graviton soft expansion~\cite{Bjerrum-Bohr:2021wwt}.

Amplitude-based methods have likewise proven to be very efficient in the context of the eikonal expansion~\cite{Collado:2018isu,KoemansCollado:2019ggb,DiVecchia:2019myk,DiVecchia:2019kta,Parra-Martinez:2020dzs,DiVecchia:2020ymx,DiVecchia:2021ndb,DiVecchia:2021bdo,Bjerrum-Bohr:2021vuf,Bjerrum-Bohr:2021din,DiVecchia:2022owy,DiVecchia:2022piu,Cristofoli:2020uzm,Bellazzini:2022wzv}. This formalism relies on the remarkable exponentiation of scattering amplitudes in impact parameter space, and it also provides the classical scattering angle from the loop expansion of amplitudes. Unitarity lies underneath the phenomenon of exponentiation and at lowest orders this is
the same mechanism that removes superclassical terms from the classical potential by means of Born subtractions.

There are thus numerous ways to extract classical general relativity
from the quantum mechanical scattering amplitudes. To the lowest orders the various 
methods are in direct correspondence with each other but as the order of perturbation theory grows they become quite different in details. Efficiency of computation thus becomes a crucial
criterion, but there are also conceptual issues related to how gravitational radiation is taken into account.

Parallel to these developments based on scattering amplitudes in a quantum field theoretic formulation of gravity, there has also been
impressive progress based on a Post-Minkowksian formulation of gravity using worldline formulations~\cite{Goldberger:2004jt,Goldberger:2006bd,Goldberger:2005cd,Kalin:2020mvi,Kalin:2020fhe,Kalin:2020lmz,Mogull:2020sak,Jakobsen:2021smu,Mougiakakos:2021ckm,Riva:2021vnj,Riva:2022fru,Liu:2021zxr,Dlapa:2021npj,Jakobsen:2021lvp,Jakobsen:2021zvh,Dlapa:2021vgp,Jakobsen:2022fcj,Jakobsen:2022psy,Kalin:2022hph,Dlapa:2022lmu,Jakobsen:2022zsx,Jakobsen:2023ndj}. In fact, at present the only full computation to fourth Post-Minkowskian order has been presented using this formalism~\cite{Dlapa:2022lmu,Dlapa:2023hsl}. These worldline formulations manifestly bypass the need to consider cancellations of superclassical terms. Computations using worldlines resemble the integration techniques for loop calculations of scattering amplitudes (a perhaps inevitable situation since both boil down to Green function methods for classical gravity), and there are indeed clear links
between worldline diagrams and scattering amplitude Feynman diagrams after localizing the loops on velocity cuts~\cite{Bjerrum-Bohr:2021din}.

One particularly important aspect of the Post-Minkowskian worldline formulation is its adaptability to scattering in a dissipative 
setting such as it appears when gravitational radiation (and back-reaction) is taken into account~\cite{Jakobsen:2022psy,Kalin:2022hph}. This requires the doubling of variables in a formulation that has its roots in the Keldysh-Schwinger closed-time path integral, making it unavoidable to work with either retarded or advanced propagators. This makes the link to conventional Feynman diagrams and
scattering matrices less obvious and concerns have indeed been raised whether conventional scattering matrices can capture
all features of gravitational radiation and the associated radiation reaction. However, the KMOC formalism expresses the change of
any physical observable during a scattering process. Although it manifestly rewrites this in terms of conventional scattering 
matrices and ordinary Feynman propagators, it should also capture all effects due to radiation. In addition, when taking the classical
limit of the KMOC formalism one essentially restricts the incoming massive states to lie on classical paths at $t = -\infty$. This
seems to closely parallel the starting point of worldline calculations. One would therefore expect that these two apparently different
approaches should be closely related. We shall here show that the two are indeed equivalent in the
classical limit. The fact that
the Keldysh-Schwinger closed time paths are used in the worldline calculations while the KMOC formalism rewrites the same observable in terms of standard Feynman
diagrams with ordinary Feynman propagators is just a convenient choice of representation.
As expected, it turns out to be advantageous to not employ the
conventional Born expansion of the $S$-matrix in terms of the scattering matrix $T$ but rather make use of the fact that the
full $S$-matrix serves as the time evolution operator. The KMOC expectation values can then be re-expressed in a path integral
form that encapsulates both forward and backward evolution in time. This is the origin of the Keldysh-Schwinger doubling of
degrees of freedom and the associated closed time paths, as expected from other contexts~\cite{Caron-Huot:2007cma,Chou:1984es}.
Once the equivalence has been established, it is also of interest to look at how the diagrammatic equivalence works out in detail. We
illustrate the correspondence up to second Post-Minkowskian order where complications from iterations (in worldline language)
and loops (in KMOC language) are already present, and from which we therefore can learn how the diagrammatic match occurs.

Specifically, in section~\ref{sec:KMOCtoObs} we show to all orders in perturbation theory that classical observables computed in the KMOC framework are identical to those obtained from the in-in worldline theory. We conclude that the specification of initial conditions in KMOC, which are formally implemented in terms of suitable wave functions, leads to the intuitively obvious initial conditions imposed in the worldline language (and when solving the classical equations of motion). The momentum kick will serve as the default example of interest. In classical mechanics or, alternatively, the worldline theory, the momentum is traditionally defined in terms of the canonical momentum conjugate to a worldline position coordinate. In contrast, in the amplitude framework the momentum operator is ultimately defined as the generator of translation of a quantum field. Thus, these constitute two quite different starting points for computing what is supposedly the same quantity. Nevertheless, we conclude that the former definition indeed emerges out of the latter in the $\hbar\to 0$-limit. 

Our work is related to a broad and rich literature on the formulation of various aspects of quantum field theory in terms of first-quantized worldlines (see, $e.g.$,~\cite{Strassler:1992,Schubert:2001,Laenen:2008,White:2011,Comberiati:2022ldk}). Additionally, similarities exist with analogous discussions on the emergence of the Schwinger-Keldysh contour in other contexts~\cite{Galley:2009px,Caron-Huot:2007cma,Chou:1984es}.


\section{From KMOC to worldline observables}\label{sec:KMOCtoObs}

As alluded to in the introduction, we shall in this section show that the closed time-paths of the dissipative worldline formulation~\cite{Jakobsen:2022psy,Kalin:2022hph}
can be derived from the classical limit of the KMOC-formalism~\cite{Kosower:2018adc} without additional assumptions. 

Before delving into the technicalities we will explain the origin of the equivalence in a more heuristic manner. Starting from KMOC~\cite{Kosower:2018adc}, we consider the change in 
an observable $\langle \Delta\hat{O}\rangle$ and insert a complete set of states $|\text{out}\rangle$\footnote{This 
complete set of states is denoted by $|\text{out}\rangle$ in order to facilitate the interpretation of it spanning a generic final state at $t = +\infty$.} to get successively, 
\begin{align}
\langle \Delta \hat{O} \rangle \nonumber&= \langle\text{in}|\hat{S}^{\dagger} \hat{O}\hat{S}|\text{in}\rangle
- \langle\text{in}|\hat{O}|\text{in}\rangle\\\label{Delta-heuristic}
&= \langle\text{in}|\hat{S}^{\dagger} [\hat{O},\hat{S}]|\text{in}\rangle\\\nonumber
&= \sum_{\rm out}\langle\text{out}|\hat{S}|\text{in}\rangle^*
\left(\langle\text{out}|\hat{O}\hat{S}|\text{in}\rangle - 
\langle\text{out}|\hat{S}\hat{O}|\text{in}\rangle\right)\\\nonumber
&= \sum_{\rm out}\langle\text{out}|\hat{S}|\text{in}\rangle^*\left(O_{\rm out} - O_{\rm in}\right)\langle\text{out}|\hat{S}|\text{in}\rangle ~,
\end{align}
where for illustration we have taken the {\em in} and {\em out} states to be eigenstates of the operator $\hat{O}$ with eigenvalues $O$. It is thus natural to interpret
the change $\langle \Delta\hat{O}\rangle$ as the average over two $S$-matrix elements where the states $|\text{out}\rangle$ can be viewed
as final states of a scattering process. The average is composed of one two-particle state propagating
forward in time from $t = -\infty$ to a 
general out-state $|\text{out}\rangle$ with (or without) radiation followed by the scattering process backwards in time (the conjugate
matrix element). If we write these two scattering processes in terms of path integral representations there will thus be two actions
of opposite signs, those opposite signs being due to the complex conjugation. 
Initial conditions are specified at $t=-\infty$ and the ``final state'' (which will also contain all the radiation)
appears as an intermediate state before returning to $t= -\infty$, and thus arises entirely from the insertion of a complete set of states.
This is the intuitive understanding of how the closed-time paths of dissipative dynamics are contained in the KMOC-formalism even though,
at the stage above, both scattering matrix elements can be computed from standard Feynman diagrams and Feynman propagators.\\

In the following, we shall go through these arguments in greater detail. This will also shed light on the details of how the initial conditions are inherited from KMOC in the worldline formalism.

Our starting point is two massive scalars in an {\em in}-state without radiation,
\begin{equation}\label{KMOC-in-state}
    \left.|\text{in}\right\rangle = \prod_{j=1,2}\int{d{\mathrm{\Pi}}_{p_j}}\tilde{\Phi}_j(\vec{p}_j)e^{\frac{i}{\hbar}b_jp_j}\left.|p_1p_2;0\right\rangle
\end{equation}
where
\beq\label{dPi}
d{\mathrm{\Pi}}_{p_j} ~\equiv~ \frac{d^{3}p_j}{(2\pi)^{3}2E_j(p_j)}
\eeq
and  $\tilde{\Phi}_j(\vec{p}_j)e^{\frac{i}{\hbar}b_jp_j}$ is the wavefunction of particle $j$ and $E_j(p_j)=\sqrt{m_j^2+p_j^2}$. We refer to ref.~\cite{Kosower:2018adc} for further details. 
We restrict our observable $\hat{O}$ so that it can be written in the single-excitation momentum basis of the scalar fields and some basis $|K\rangle$ for the gravitational field as follows
\begin{equation}\label{dO-general}
    \hat{O}=\hat{O}^{\dagger}=\sum_K \int d\mathrm{\Pi}_{p_1}d\mathrm{\Pi}_{p_2}|{p_1p_2;K}\rangle O(\vec{p},i\hbar\partial_{\vec{p}};K)\langle p_1p_2;K| ~.
\end{equation}
We use a compact notation that suppresses scalar particle labels in the argument of $O$ because indices will soon proliferate.
The symbolic sum over $K$ is a short-hand for the full on-shell phase space integrations and polarization sums of an arbitrarily large number of gravitons, or, alternatively, the functional integral over metric configurations, depending on which basis is most convenient. Eventually we will be most interested in the special case where $O$ is taken to be the momentum of one of the scalar particles $O(\vec{p},i\hbar\partial_{\vec{p}};K)=p^\mu_1$, but initially we will be more general. Other interesting observables which also depend on $i\hbar\partial_{\vec{p}}$ could for example be the angular momentum operator.

Let us now express the complete set of states $|\text{out}\rangle$ in eq.~(\ref{Delta-heuristic}) as a sum over single particle excitations of the scalar fields and any number of gravitons 
\beq
1 =  \sum_K \int d\mathrm{\Pi}_{l_1}d\mathrm{\Pi}_{l_2}|l_1l_2;K \rangle\langle l_1l_2;K| ~.
\eeq
Using eqs.~(\ref{KMOC-in-state}) and~(\ref{dO-general}), we can then write~(\ref{Delta-heuristic}) as
\begin{align}\label{dO-Smatrix}
    \langle\Delta \hat{O}\rangle = \sum_K&\prod_{j=1,2}\int{d{\mathrm{\Pi}}_{p_j}d{\mathrm{\Pi}}_{p'_j}d{\mathrm{\Pi}}_{l_j}}\tilde{\mathrm{\Phi}}_j{(\vec{p}_j{'})}^{*}\tilde{\mathrm{\Phi}}_j(\vec{p}_j) e^{\frac{i}{\hbar}b_j(p_j-p_j')}\langle l_1l_2;K|\hat{S}|p'_1p'_2;0\rangle^{*}\\\nonumber &\times \left[O(\vec{l},i\hbar\partial_{\vec{l}};K)- O(\vec{p},-i\hbar\partial_{\vec{p}};0)\right]\langle l_1l_2;K|\hat{S}|p_1p_2;0\rangle.
\end{align}
The minus sign on the argument, $-i\hbar\partial_{\vec{p}}$, in the second term arises due to conjugation and the assumed Hermiticity of the observable in (\ref{dO-general}), $i.e.$,
\begin{align}
    \langle l_1l_2;K|\hat{S}\hat{O}|p_1p_2;0\rangle &= \langle p_1p_2;0|\hat{O}\hat{S}^{\dagger}|l_1l_2;K\rangle^{*}\nonumber\\\newline\nonumber
    &= O(\vec{p},i\hbar\partial_{\vec{p}})^*\langle p_1p_2;0|\hat{S}^{\dagger}|l_1l_2;K\rangle^{*}\\\newline\nonumber
    &= O(\vec{p},-i\hbar\partial_{\vec{p}})\langle l_1l_2;K|\hat{S}|p_1p_2;0\rangle.\nonumber
\end{align}

Our aim is to write the $S$-matrix elements in terms of path integrals. First we introduce the scalar field operator
\begin{equation}\label{phi-operator}
     \hat{\phi}_j(t,\vec{x})=\frac{1}{\hbar^{3/2}}\int \frac{d^3p}{(2\pi)^3\sqrt{2E_j(p)}}\left\{\hat{a}_{jp}(t)e^{-\frac{i}{\hbar}[E_{j}(p)(t-t_\text{in})-\vec{p}\cdot\vec{x}]}+\text{h.c.}\right\},
\end{equation}
where $t_{\text{in}}$ is the initial time, which we will push to $-\infty$. The phase containing $t_{\text{in}}$ says that we have defined the creation and annihilation operators such that at time $t=t_\text{in}$ the interaction-picture operator $\hat{a}_{jp}(t_\text{in})$ coincides with the Heisenberg-picture operator. The scalar field momentum 
eigenstates can be written in terms of the $a_{jp}$-operators appearing on the right-hand side of~(\ref{phi-operator}) in the usual way
\begin{equation}\label{p1p2}
    |p_1p_2\rangle \equiv \sqrt{2E_1(p_1)2E_2(p_2)}\hat{a}^{\dagger}_{1p_1}(t_{\text{in}})\hat{a}^{\dagger}_{2p_2}(t_\text{in})|0\rangle
\end{equation}

We now define the initial eigenstates $\left|\phi_j\right\rangle$ of the field operator~(\ref{phi-operator}), 
\begin{equation}\label{phi-eigenstate}
\hat{\phi}_j(t_{\text{in}},\vec{x})\left|\phi_j\right\rangle=\phi_j(\vec{x})\left|\phi_j\right\rangle
\end{equation}
and likewise for the gravitational analog of~(\ref{phi-operator})
\begin{equation}\label{h-eigenstate}
    \hat{h}_{\mu\nu}(t_{\text{in}},\vec{x})\left|h\right\rangle=h_{\mu\nu}(\vec{x})\left|h\right\rangle
\end{equation}
We can write the $S$-matrix in the basis of~(\ref{h-eigenstate}) as follows
\begin{equation}\label{S-pathintegral}
    \hat{S}=\hat{S}(t_\text{in},t_\text{f})=\int Dh e^{\frac{i}{\hbar}S_h[h]}\left|h_{\text{f}}\right\rangle \hat{S}_\phi(h)\left\langle h_\text{in}\right|
\end{equation}
where
\beq
Dh\equiv \mathcal{N}\prod_{\vec{x}\,\rho\lambda}\prod_{t_\text{in}\leq t\leq t_\text{f}} dh_{\rho\lambda}(t,\vec{x})
\eeq
 and 
 \beq
 S_h[h]\equiv \int_{t_\text{in}}^{t_\text{f}}dt\int d^3x \sqrt{-g}\mathcal{L}_{EH}+S_{GF}[h]
 \eeq
 is the gauge-fixed Einstein-Hilbert action and $|h_\text{f}\rangle$ is the eigenstate of $\hat{h}_{\mu\nu}(t_{\text{in}},\vec{x})$ corresponding to the eigenvalue $h_{\mu\nu}(t_\text{f},\vec{x})$ while $|h_\text{in}\rangle$ is the eigenstate corresponding to the eigenvalue $h_{\mu\nu}(t_\text{in},\vec{x})$. We will deal with the normalization $\mathcal{N}$ in the very end when everything is put together. The operator $\hat{S}_\phi(h)$ in~(\ref{S-pathintegral}) is defined as
\begin{equation}\label{Sphi-pathintegral}
    \hat{S}_\phi(h)=\hat{S}_{\phi 1}(h)\hat{S}_{\phi 2}(h)=\prod_{j=1,2}\int D\phi_j e^{\frac{i}{\hbar}S_{\phi j}[h,\phi_j]}\left|\phi_{j,\text{f}}\right\rangle \left\langle \phi_{j,\text{in}}\right|
\end{equation}
where the notation in~(\ref{Sphi-pathintegral}) is analogous to~(\ref{S-pathintegral}). The $S_{\phi j}[h,\phi_j]$ of eq.~(\ref{Sphi-pathintegral}) is the action of a scalar field in curved 
space-time (see e.g., the last term in eq. (\ref{full-action}) below).\\

We can now rewrite the S-matrix elements appearing in~(\ref{dO-Smatrix}) using eq.~(\ref{S-pathintegral})
\begin{equation}\label{S-matrix}
    \langle l_1l_2;K|\hat{S}|p_1p_2;0\rangle=\int Dh e^{\frac{i}{\hbar}S_h[h]} \langle K|h_\text{f}\rangle\langle h_\text{in}|0\rangle\langle l_1l_2|\hat{S}_\phi(h)|p_1p_2\rangle
\end{equation}
This allows us to focus on the $\hat{S}_\phi(h)$-matrix elements with $h_{\mu\nu}$ playing the role of external sources. We can rewrite the $\hat{S}_\phi(h)$-matrix elements in terms of the field operators defined in~(\ref{phi-operator}) acting on the vacuum as follows
\begin{align}\label{Sphi-matrix}
    \langle l_1l_2|\hat{S}_\phi(h)|p_1p_2\rangle=\prod_{j=1,2}&\frac{2E_j(l_j)}{\hbar^{3/2}}\frac{2E_j(p_j)}{\hbar^{3/2}}\int d^3y_j d^3x_j e^{-\frac{i}{\hbar}\vec{l}_j\cdot\vec{y}_j+\frac{i}{\hbar}\vec{p}_j\cdot\vec{x}_j}\\\nonumber&\times\langle 0|\hat{\phi}_j(t_\text{in},\vec{y}_j)\hat{S}_{\phi j}(h)\hat{\phi}_j(t_\text{in},\vec{x}_j)|0\rangle
\end{align}
We next consider the quantity
$$
\langle 0|\hat{\phi}_j(t_\text{in},\vec{y}_j)\hat{S}_{\phi j}(h)\hat{\phi}_j(t_\text{in},\vec{x}_j)|0\rangle
$$
in eq.~(\ref{Sphi-matrix}). With the aim to make a transition to a worldline formulation we include a source $J$ for the scalar field $\phi$ in the action,
\beq
S_{\phi j}[\phi_j,h] \to S_{\phi j}[\phi_j,h,J] \equiv  S_{\phi j}[\phi_j,h] + \int d^4x \sqrt{-g}J(x)\phi_j(x)
\eeq
so that the eigenvalues of the scalar field operators can be represented by variations with respect to the source at the temporal end points. The generating functional is defined as
\begin{eqnarray}\label{generating-functional}
    Z_j\left[J;h\right] &\equiv& \int{D\phi }e^{\frac{i}{\hbar }S_{\phi j}[\phi_j,h,J]}
\langle 0|{\phi }_{\mathrm{f},j}\rangle \langle {\phi }_{\mathrm{in},j}|0\rangle  \cr
&=& Z_j\left[0;h\right]e^{-\frac{i}{2\hbar}\int d^4xd^4y\sqrt{-g\left(y\right)}J(y)\mathrm{\Delta }_j\left(x,y;h\right)\sqrt{-g\left(x\right)}J(x)}
\end{eqnarray}
where the Green function $\mathrm{\Delta }_j\left(x,y;h\right)$ is defined as the solution to 
\beq\label{green}
    -\left(\frac{1}{\sqrt{-g}}{\partial }_{\mu }(\sqrt{-g}g^{\mu \nu }{\partial }_{\nu })+\frac{m^2_j}{{\hbar }^2}-i\epsilon\right)\mathrm{\Delta }_j\left(x,y;h\right)=\frac{{\delta }^4\left(x-y\right)}{\sqrt{-g}} ~.
\eeq
The shown $i\epsilon$-prescription is a consequence of the presence of the factor 
\begin{equation}
\langle 0|{\phi }_{\mathrm{f},j}\rangle \langle {\phi }_{\mathrm{in},j}|0\rangle\sim \lim_{\epsilon\to 0^+}\text{exp}\left(-\frac{1}{2}\epsilon\int d^4x\phi_j^2\right)
\end{equation}
in eq.~(\ref{generating-functional}) (see, {\em e.g.}, ref.~\cite{Weinberg:1995mt} for details). If we insert $\hat{S}_{\phi j}(h)$ given in~(\ref{Sphi-pathintegral}) into $\langle 0|\hat{\phi}_j(t_\text{in},\vec{y}_j)\hat{S}_{\phi j}(h)\hat{\phi}_j(t_\text{in},\vec{x}_j)|0\rangle$ and write the result in terms of variations of~(\ref{generating-functional}), we obtain:
\begin{align}\label{phiS(h)phi}
    \langle 0|\hat{\phi }_j(t_{\mathrm{in}},\vec{y}_j){\hat{S}}_{\phi j}(h)\hat{\phi}_j(t_{\mathrm{in}},\vec{x}_j)|0\rangle &=i\hbar \mathrm{\Delta }_j\left(\left(t_{\mathrm{in}},\vec{x}_j\right),\left(t_{\mathrm{f}},\vec{y}_j\right);h\right)Z_j\left[0;h\right]\\\nonumber&=i\hbar \mathrm{\Delta }_j\left(\left(t_{\mathrm{in}},\vec{x}_j\right),\left(t_{\mathrm{f}},\vec{y}_j\right);h\right)\langle 0|{\hat{S}}_{\phi j}(h)|0\rangle ~.
\end{align}
 Combining terms, we find:

\begin{equation}\label{S-matrix-G}
    \langle l_1l_2;K|\hat{S}|p_1p_2;0\rangle=\prod_{j=1,2}4E_j(p_j)E_j(l_j)\int d^3y_jd^3x_j e^{\frac{i}{\hbar}(\vec{p}_j\cdot\vec{x}_j-\vec{l}_j\cdot\vec{y}_j)} G_j(\vec{x}_j,\vec{y}_j;0,K),
\end{equation}
where
\begin{equation}\label{G-def}
    G_j(\vec{x}_j,\vec{y}_j;K_0,K)\equiv \frac{i}{\hbar^2}\int Dh e^{\frac{i}{\hbar}S_h[h]}\langle K|h_\text{f}\rangle\langle h_\text{in}|K_0\rangle\langle 0|\hat{S}_{\phi j}(h)|0\rangle\Delta_j((t_\text{in},\vec{x}_j),(t_\text{f},\vec{y}_j);h)
\end{equation}
In this expression, the factor 
\begin{equation}
\langle 0|\hat{S}_{\phi j}(h)|0\rangle=\int{D\phi_j}e^{\frac{i}{\hbar }S_{\phi j}[\phi_j,h]}
\langle 0|{\phi }_{\mathrm{f},j}\rangle \langle {\phi }_{\mathrm{in},j}|0\rangle
\end{equation}
represents fluctuations of the scalar fields, which can be ignored in the classical limit (the number of scalar matter particles is assumed to be fixed at all times). Up to a proportionality factor, one may
interpret $G_j(\vec{x}_j,\vec{y}_j;K_0,K)$ in the classical limit as the scalar Green function $\Delta_j((t_\text{in},\vec{x}_j),(t_\text{f},\vec{y}_j);h_\text{cl})$ in a classical gravitational background $h_\text{cl}$ determined by the boundary conditions $K$ and $K_0$. 

These definitions allow us to write the original observable $\langle\Delta \hat{O}\rangle $ back in (\ref{dO-Smatrix}) in the following form. Note that the various factors of $2E_j(p)$ in (\ref{dPi}) and (\ref{S-matrix-G}) cancel, except for one case

\begin{align}\label{dO-xp}
    \langle\Delta \hat{O}\rangle = \sum_K&\prod_{j=1}^2\int d^3y_jd^3x_jd^3y'_jd^3x'_j G_j(\vec{x}_j,\vec{y}_j;0,K)G_j(\vec{x}_j{'},\vec{y}_j{'};0,K)^*\\\nonumber
    &\times\int \frac{d^3p_jd^3p'_jd^3l_j}{(2\pi)^9}\tilde{\Phi}_j(\vec{p}_j)\tilde{\Phi}_j(\vec{p}_j{'})^*e^{-\frac{i}{\hbar}\vec{b}_j\cdot(\vec{p}_j-\vec{p}_j{'})}e^{\frac{i}{\hbar}(\vec{l}_j\cdot\vec{y}_j{'}-\vec{p}_j{'}\cdot\vec{x}_j{'})}\\\nonumber
    &\times\left[O(\vec{l},i\hbar \partial_{\vec{l}};K)-O(\vec{p},-i\hbar \partial_{\vec{p}};0)\right]2E_j(l_j)e^{-\frac{i}{\hbar}(\vec{l}_j\cdot\vec{y}-\vec{p}_j\cdot\vec{x}_j)}
\end{align}
We have here chosen a frame where $b^0_j=0$ for simplicity. To move further towards a resemblance with the worldline formalism, we will take a closer look at the momentum integrals. We first consider the last line of eq.~(\ref{dO-xp}). The derivative $i\hbar\partial_{l_j}$ in the operator $O(\vec{l},i\hbar \partial_{\vec{l}};K)$ acts
as follows
\begin{align}
    i\hbar\partial_{\vec{l}_j}\left[E_j(l_j)e^{-\frac{i}{\hbar}(\vec{l}_j\cdot\vec{y}-\vec{p}_j\cdot\vec{x})}\right]&=\left[\vec{y}+i\hbar\partial_{l_j}\text{log}E_j(l_j)\right]E_j(l_j)e^{-\frac{i}{\hbar}(\vec{l}_j\cdot\vec{y}-\vec{p}_j\cdot\vec{x})}\\\nonumber
    &=\left[\vec{y}+\mathcal{O}(\hbar)\right]E_j(l_j)e^{-\frac{i}{\hbar}(\vec{l}_j\cdot\vec{y}-\vec{p}_j\cdot\vec{x})}
\end{align}
As long as we are only interested in the classical limit, we can therefore replace $i\hbar\partial_{\vec{l}_j}\to \vec{y}_j$ in~(\ref{dO-xp}). Additionally, we can replace $-i\hbar\partial_{\vec{p}_j}\to \vec{x}_j$ and also $\vec{p}_j\to -i\hbar\partial_{\vec{x}_j}$ and $\vec{l}_j\to i\hbar\partial_{\vec{y}_j}$. This allows us to write eq.~(\ref{dO-xp}) as follows

\begin{align}\label{dO-G}
    \langle\Delta \hat{O}\rangle = \sum_K\prod_{j=1}^2\int &d^3y_jd^3x_jd^3y'_jd^3x'_j G_j(\vec{x}_j,\vec{y}_j;0,K)\\\nonumber
    &\times \left\{O[i\hbar \partial_{\vec{y}},\vec{y};K]-O[-i\hbar \partial_{\vec{x}},\vec{x};0]\right\}\\\nonumber
    &\times \mathcal{E}_j(\vec{y}_j-\vec{y}_j{'})\Phi_j(\vec{x}_j)\Phi_j(\vec{x}_j{'})^*G_j(\vec{x}_j{'},\vec{y}_j{'};0,K)^*
\end{align}
where
\beq\label{E(y-y')}
    \mathcal{E}_j(\vec{y}_j-\vec{y}_j{'}) \equiv\int \frac{d^3l_j}{(2\pi)^3}2E_j(l_j)e^{\frac{i}{\hbar}\vec{l}_j(\vec{y}_j-\vec{y}_j{'})}
    =2\hbar^3E_j(i\hbar\partial_{\vec{y}_j})\delta^3(\vec{y}_j-\vec{y}_j{'}),
\eeq
and where
\begin{equation}\label{Phi(x)}
   \Phi_j(\vec{x}_j)\equiv \int \frac{d^3p}{(2\pi)^3}\tilde{\Phi}_j(\vec{p})e^{\frac{i}{\hbar}\vec{p}\cdot(\vec{x}_j-\vec{b}_j)}.
\end{equation}
are the Fourier transform of the wave functions.\\

We now focus on the integrals over the primed variables $\vec{x}_j{'},\vec{y}_j{'}$ in eq.~(\ref{dO-G}). The $\vec{y}_j{'}$-integral is readily done by exploiting the $\delta$-function in~(\ref{E(y-y')}). We deal with the $\vec{x}_j{'}$-integral by moving it to the last couple of factors in~(\ref{dO-G}) that depend on $x_j{'}$ (which we will temporarily denote $I$) and undo the Fourier transform of the wave functions to get
\begin{align}
I&\equiv
\int d^3x'_j\Phi_j(\vec{x}_j)\Phi_j(\vec{x}_j{'})^*G_j(\vec{x}_j{'},\vec{y}_j{'};0,K)^*\\
&=\hbar^3\int d^3x'_j\int \frac{d^3pd^3q}{(2\pi)^6}\tilde{\Phi}_j(\vec{p})^*\tilde{\Phi}_j(\vec{p}+\hbar\vec{q})e^{-\frac{i}{\hbar}\vec{p}\cdot(\vec{x}_j{'}-\vec{b}_j)+\frac{i}{\hbar}(\vec{p}+\hbar\vec{q})\cdot(\vec{x}_j-\vec{b}_j)}G_j(\vec{x}_j{'},\vec{y}_j{'};0,K)^*.\nonumber
\end{align}
In the last line, we have made the change of variables $\vec{p}{\ '}=\vec{p}+\hbar\vec{q}$, where the factor of $\hbar$ is a question of convenience at present.
Using the same line of reasoning as in ref.~\cite{Kosower:2018adc} we now replace $\tilde{\Phi}_j(\vec{p})^*\tilde{\Phi}_j(\vec{p}+\hbar\vec{q})=|\tilde{\Phi}_j(\vec{p})|^2+\mathcal{O}(\hbar)$. If we further make a change of variables $\vec{x}_j{'}=\vec{x}_j+\Delta \vec{x}_j$, we get, successively,
\begin{align}\label{I}
I&= \hbar^3\delta^3(\vec{x}_j-\vec{b}_j)\int \frac{d^3p}{(2\pi)^3}|\tilde{\Phi}_j(\vec{p})|^2\int d^3\Delta x_j e^{-\frac{i}{\hbar}\vec{p}\cdot\Delta\vec{x}_j}G_j(\vec{x}_j+\Delta \vec{x}_j,\vec{y}_j{'};0,K)^*\\\nonumber
&= \hbar^3\delta^3(\vec{x}_j-\vec{b}_j)\int \frac{d^3p}{(2\pi)^3}|\tilde{\Phi}_j(\vec{p})|^2\int d^3\Delta x_j e^{-\frac{i}{\hbar}\Delta\vec{x}_j\cdot (\vec{p}+i\hbar\partial_{\vec{x}_j})}G_j(\vec{x}_j,\vec{y}_j{'};0,K)^*\\\nonumber
&= \hbar^6\delta^3(\vec{x}_j-\vec{b}_j)\int d^3p|\tilde{\Phi}_j(\vec{p})|^2\delta^3(\vec{p}+i\hbar\partial_{\vec{x}_j})G_j(\vec{x}_j,\vec{y}_j{'};0,K)^*\\\nonumber
&= \hbar^6\delta^3(\vec{x}_j-\vec{b}_j)|\tilde{\Phi}_j(-i\hbar\partial_{\vec{x}_j})|^2G_j(\vec{x}_j,\vec{y}_j{'};0,K)^*
\end{align}
When we next insert this into eq.~(\ref{dO-G}), we can immediately perform the $x_j$-integral using the $\delta$-function $\delta^3(\vec{x}_j-\vec{b}_j)$. Furthermore, we can also perform the $y_j{'}$ using the $\delta$-function in the second equality of eq.~(\ref{E(y-y')}). This leaves us with only the $y_j$-integral
\begin{align}\label{dO-G-v2}
    \langle\Delta \hat{O}\rangle = 2\hbar^9\sum_K&\prod_{j=1}^2\int d^3y_j G_j(\vec{b}_j,\vec{y}_j;0,K)\\\nonumber&\times\left[O(i\hbar \partial_{\vec{y}},\vec{y};K)-O(-i\hbar \partial_{\vec{b}},\vec{b};0)\right]\\\nonumber
    &\times E_j(i\hbar\partial_{\vec{y}_j})|\tilde{\Phi}_j(-i\hbar\partial_{\vec{b}_j})|^2G_j(\vec{b}_j,\vec{y}_j;0,K)^*
\end{align}
Having obtained the result in~(\ref{dO-G-v2}), we now go back and unpack the expression for $G_j$ given in eq.~(\ref{G-def}); in particular, we want to write the Green function $\mathrm{\Delta }_j((t_{\mathrm{in}},\vec{b}_j),(t_{\mathrm{f}},\vec{y}_j);h)$ defined by eq.~(\ref{G-def}) in terms of its worldline-representation~\cite{Bastianelli:2002fv,Bastianelli:1998jm}
\begin{equation}\label{Schwinger-feynman}
    \mathrm{\Delta }_j\left(x,y;h\right)=\mathcal{N}\int^{\infty }_{0}{d T }\int^{z\left( T \right)=y}_{z\left(0\right)=x}{\mathcal{D}z}e^{-\frac{i}{\hbar }\frac{m_j}{2}\int^{ T }_{0}{ds}\left\{g_{\mu \nu }\left(z\right){\dot{z}}^{\mu }{\dot{z}}^{\nu }-\frac{{\hbar }^2}{4m_j^2}R(z)+1\right\}},
\end{equation}
where $\mathcal{D}z\equiv \prod_{0<\tau<T}d^4z(\tau)\sqrt{-g(z(\tau))}$, $\mathcal{N}$ is a constant, and $R$ is the Ricci scalar. Equation~(\ref{Schwinger-feynman}) includes all quantum effects; this is overkill for our purpose, so we first make some approximations that hold exactly in the classical limit. First, we note that the $T$-integral in~(\ref{Schwinger-feynman}) can be done in the classical limit using the principle of stationary phase (see appendix~\ref{classical-worldline-representation}). It is interesting to note that the same result was reached in ref.~\cite{Mogull:2020sak} without invoking the principle of stationary phase, but instead by using the on-shell initial conditions. With a few more simplifications that are also shown in detail in the appendix and which only consist in removing quantum mechanical corrections, we can effectively replace~(\ref{Schwinger-feynman}), as evaluated with the relevant boundaries, by
\begin{equation}\label{Schwinger-feynman-classical}
    \mathrm{\Delta }_j((t_{\mathrm{in}},\vec{b}_j),(t_{\mathrm{f}},\vec{y}_j);h)= \mathcal{N}\int^{z_j({\tau }_{\mathrm{f},j})=(t_{\mathrm{f}},\vec{y}_j)}_{z_j({\tau }_{\mathrm{in},j})=\left(t_{\mathrm{in}},{\vec{b}}_{j}\right)}{Dz_j}e^{\frac{i}{\hbar }\int^{{\tau }_{\mathrm{f},j}}_{{\tau }_{\mathrm{in},j}}{d\tau }L_j\left[z_j,\dot{z}_j,h\right]}
 \end{equation}
which is valid in the classical limit (here, some factors have been absorbed into $\mathcal{N}$). The Lagrangian appearing in (\ref{Schwinger-feynman-classical}) is the Polyakov form of the classical Lagrangian for a point particle on curved space-time:
\beq\label{polyakov-lagrangian}
L_j\left[z_j,\dot{z}_j,h\right]=-\frac{m_j}{2}\left[g_{\mu \nu }\left(z_j\right)\frac{dz^{\mu }_j}{d\tau }\frac{dz^{\nu }_j}{d\tau }+1\right],
\eeq
The derivation of~(\ref{Schwinger-feynman-classical}) in appendix~\ref{classical-worldline-representation} also shows that we should identify the length of the parameter interval $\tau_{\text{f},j}-\tau_{\text{in},j}$ with the classical proper time difference between the space-time points $(t_{\mathrm{in}},\vec{b}_j)$ and $(t_{\mathrm{f}},\vec{y}_j)$ that get pushed to infinity when we push $t_{\mathrm{f}}-t_{\mathrm{in}}$ to infinity. 

We now insert~(\ref{Schwinger-feynman-classical}) into eq.~(\ref{G-def}). Although straightforward, the expression we obtained in eq.~(\ref{dO-G-v2}), also contains derivatives such as
 $$
i\hbar\partial_{\vec{y}_j}G_j(\vec{b}_j,\vec{y}_j;0,K)^*
$$
which we need to consider carefully. These derivatives will act on  $\mathrm{\Delta }_j((t_{\mathrm{in}},\vec{b}_j),(t_{\mathrm{f}},\vec{y}_j);h)^*$ so we have to evaluate derivatives of the boundaries of the worldline path-integral in~(\ref{Schwinger-feynman-classical}) which is slightly non-trivial. The full evaluation of these relations is provided in appendix~\ref{derivative-worldline}. The result is
\begin{align}\label{derivative-of-propagator}
    {\left(-i\hbar {\partial }_y\right)}^n{\mathrm{\Delta }}_j\left(x,y;h\right)^* &= \mathcal{N}\int^{z({\tau }_{\mathrm{f},j})=y}_{z({\tau }_{\mathrm{in},j})=x}{Dz}\left\{{\left(-{\left.\frac{\partial L_j}{\partial {\dot{z}}}\right|}_{\tau ={\tau }_{\mathrm{f},j}}\right)}^n+\mathcal{O}(\hbar )\right\}e^{-\frac{i}{\hbar }\int^{{\tau }_{\mathrm{f},j}}_{{\tau }_{\mathrm{in},j}}{d\tau }L_j}\cr
{\left(-i\hbar {\partial }_x\right)}^n{\mathrm{\Delta }}_j\left(x,y;h\right)^* &= \mathcal{N}\int^{z({\tau }_{\mathrm{f},j})=y}_{z({\tau }_{\mathrm{in},j})=x}{Dz}\left\{{\left({\left.\frac{\partial L_j}{\partial {\dot{z}}}\right|}_{\tau ={\tau }_{\mathrm{in},j}}\right)}^n+\mathcal{O}(\hbar )\right\}e^{-\frac{i}{\hbar }\int^{{\tau }_{\mathrm{f},j}}_{{\tau }_{\mathrm{in},j}}{d\tau }L_j}.
 \end{align}
It is convenient to introduce 
\begin{equation}\label{canonical-momentum}
    \mathcal{P}_{j}^\mu(\tau)\equiv
-\eta^{\mu\nu}\frac{\partial L_j[z_j,\dot{z}_j;h]}{\partial \dot{z}_j^{\nu}}=( \mathcal{P}_{j}^0(\tau), \vec{\mathcal{P}}_{j}(\tau))
\end{equation}
Based on the Lagrangian~(\ref{polyakov-lagrangian}), this quantity may also be written as $\mathcal{P}_{j\mu}=m_jg_{\mu\nu}(z_j)\dot{z}_j^\nu$. Substituting~(\ref{Schwinger-feynman-classical}) into eq.~(\ref{G-def}) and using eqs.~(\ref{derivative-of-propagator}) and~(\ref{canonical-momentum}), we obtain
\begin{align}\label{dO-worldline}
    \langle\Delta \hat{O}\rangle = &\mathcal{N}\sum_K \int{Dh^{(1)}Dh^{(2)}}\langle K|h^{(1)}_{\mathrm{f}}\rangle\langle h^{(1)}_{\mathrm{in}}|0\rangle\langle K|h^{(2)}_{\mathrm{f}}\rangle^{*}\langle h^{(2)}_{\mathrm{in}}|0\rangle^{*}\\\nonumber &\times \prod_{j=1}^2\int d^3y_j\int^{z_j^{(i)}({\tau }_{\mathrm{f},j})=(t_{\mathrm{f}},\vec{y}_j)}_{z_j^{(i)}({\tau }_{\mathrm{in},j})=(t_{\mathrm{in}},\vec{b}_j)}{Dz_j^{(1)}Dz_j^{(2)}}E_j(\vec{\mathcal{P}}_{j}^{(2)}(\tau_{\text{f},j}))|\tilde{\Phi}_j(\vec{\mathcal{P}}_{j}^{(2)}(\tau_{\text{in},j}))|^2
 \\\nonumber&\times\left\{O[\vec{\mathcal{P}}^{(2)}(\tau_{\text{f}}),\vec{z}^{(2)}(\tau_\text{f});K]-O[\vec{\mathcal{P}}^{(2)}(\tau_{\text{in}}),\vec{z}^{(2)}(\tau_\text{in});0]\right\}e^{\frac{i}{\hbar}(S[z_j^{\left(1\right)},h^{(1)}]-S[z_j^{\left(2\right)},h^{(2)}])}
\end{align}
The prefactor $\mathcal{N}$ in~(\ref{dO-worldline}) is a new accumulation of constants. We have also introduced the full action $S[z_j,h]$ as the combination
\beq\label{action}
    S[z_j,h] \equiv S_h\left[h\right]+\int^{{\tau }_{\mathrm{f},j}}_{{\tau }_{\mathrm{in},j}}{d\tau }L_j\left[z_j,\dot{z}_j,h\right]+\mathcal{O}(\hbar)
\eeq
where, as anticipated earlier, we have ignored the extra terms of quantum origin. These terms, in particular, can create scalar particles.

We note that $-i\hbar\partial_{\vec{b}_j}$ in~(\ref{dO-G-v2}) has how become replaced by $\vec{\mathcal{P}}^{(2)}_j(\tau_{\text{in},j})$, and $i\hbar\partial_{\vec{y}_j}$ has become 
replaced by 
$\vec{\mathcal{P}}^{(2)}_j(\tau_{\text{f},j})$. Having both $G_j$ and its complex conjugate in eq.~(\ref{dO-G-v2}), we have been forced to double the degrees of freedom by using two sets of fields $(z^{(i)},h^{(i)})$, $i = 1,2$. Here we see the emergence of the Keldysh-Schwinger path integral prescription. This is as expected since the analysis is fully general and
can include dissipation in the form of gravitational radiation.

\subsection{Restriction to single-particle observables}
The result~(\ref{dO-worldline}) is unwieldy and it may be difficult to see the forest for the trees. It can be made more clear if we give up a bit of generality and assume that the observable is restricted to one of the two scalar particles, say particle 1.  That is, we now make the following replacement
\begin{equation}\label{one-particle}
    O[\vec{\mathcal{P}}^{(2)}(\tau),\vec{z}^{(2)}(\tau);K]\to O_1[\vec{\mathcal{P}}^{(2)}_1(\tau_{1}),\vec{z}^{(2)}_1(\tau_{1})]
\end{equation}
and $O_1$ can be regarded as a function of the proper time of particle 1. This allows us to write the difference between the final and initial value in eq.~(\ref{dO-worldline}) as an integral over the proper time of particle 1
\begin{equation}\label{O-integral}
    O_1[\vec{\mathcal{P}}^{(2)}_1(\tau_{\text{f},1}),\vec{z}^{(2)}_1(\tau_{\text{f},1})]-O_1[\vec{\mathcal{P}}^{(2)}_1(\tau_{\text{in},1}),\vec{z}^{(2)}_1(\tau_{\text{in},1})]=\int_{\tau_{\text{in},1}}^{\tau_{\text{f},1}} d\tau_1 \frac{d}{d\tau_1}O_1
\end{equation}

Having assumed that $O_1$ is independent of $K$ we can remove it from~(\ref{dO-worldline}) by evaluating the sum
\begin{equation}\label{delta(h1-h2)}
    \sum_K \langle K|h^{(1)}_{\mathrm{f}}\rangle\langle K|h^{(2)}_{\mathrm{f}}\rangle^{*}=\langle h^{(2)}_{\mathrm{f}}|h^{(1)}_{\mathrm{f}}\rangle=\prod_{\mu\nu,\vec{x}'}\delta (h^{(1)}_{\mu\nu}(t_\text{f},\vec{x}')-h^{(2)}_{\mu\nu}(t_\text{f},\vec{x}'))
\end{equation}
We next introduce the following short-hand notation for the path integral over these ``closed time paths''
\begin{equation}\label{in-in-def-h}
    \int\limits_{\text{in-in}}{Dh^{(i)}}(...)\equiv  \int{Dh^{(1)}Dh^{(2)}}\langle 0|h_{\text{in}}^{(2)}\rangle\langle h_{\text{in}}^{(1)}|0\rangle\prod_{\mu\nu,\vec{x}'}\delta\left(h^{(1)}_{\mu\nu}(t_\text{f},\vec{x}')-h^{(2)}_{\mu\nu}(t_\text{f},\vec{x}')\right)(...)
\end{equation}
With this notation and with the stated restriction of the observable in~(\ref{one-particle}), we can now write eq.~(\ref{dO-worldline}) more compactly as
\begin{align}\label{dO-worldline-v2}
    \langle\Delta \hat{O}\rangle = &\mathcal{N} \int\limits_{\text{in-in}}{Dh^{(i)}} \int d^3y_j\int^{z_j^{(i)}({\tau }_{\mathrm{f},j})=\left(t_{\mathrm{f}},\vec{y}_j\right)}_{z_j^{(i)}({\tau }_{\mathrm{in},j})=\left(t_{\mathrm{in}},\vec{b}_j\right)}{Dz_j^{(i)}}
 |\tilde{\Phi}_{j}(\vec{\mathcal{P}}_{j,\text{in}}^{(2)})|^2\\\nonumber&\times \left(E_j(\vec{\mathcal{P}}_{j,\text{f}}^{(2)})\int_{\tau_{\text{in},1}}^{\tau_{\text{f},1}} d\tau_1\frac{d}{d\tau_1}O_1[\vec{\mathcal{P}}^{(2)}_1,\vec{z}^{(2)}_1]\right)e^{\frac{i}{\hbar}(S[z_j^{\left(1\right)},h^{(1)}]-S[z_j^{\left(2\right)},h^{(2)}])}
\end{align}
which implicitly includes the obvious product over particle labels $j=1,2$ and over the two copies of variables $i=1,2$.   The KMOC difference in observables from $t=-\infty$ to $t=+\infty$ has thus been expressed as (the time integral of) a one-point function over paths as in the Keldysh-Schwinger prescription.

\subsubsection{Interpretation of boundary/initial conditions}
Having obtained~(\ref{dO-worldline-v2}), we would like to understand the consequences of having the (squared) wave function $|\tilde{\Phi}_{j}(\vec{\mathcal{P}}_{j,\text{in}}^{(2)})|^2=|\tilde{\Phi}_j(\vec{\mathcal{P}}_{j}^{(2)}(\tau_{\text{in},j}))|^2$  inside the path integral. In line with KMOC~\cite{Kosower:2018adc}, we assume that for widely separated scalar particles, the function $|\tilde{\Phi}_j(\vec{p})|^2$ will be strongly centered around the classical momentum $\vec{p}\sim m_j\vec{u}_j$; $i.e.$
\begin{equation}\label{delta(mu-P)}
    |\tilde{\Phi}(\vec{\mathcal{P}}_{j,\text{in}}^{(2)})|^2\approx 2E_j(m_j\vec{u}_j)\delta^3(m_j\vec{u}_j-\vec{\mathcal{P}}^{(2)}_{j,\text{in}})
\end{equation}
The interpretation of the presence of $|\tilde{\Phi}(\vec{\mathcal{P}}^{(2)}_{j,\text{in}})|^2$ in the path integral~(\ref{dO-worldline-v2}) is then clear if we view the time-discretized version of the path integral, and replace
\begin{align}\label{P-discrete}
    \mathcal{P}^{(2)}_{j\mu,\text{in}}&=m_j\eta_{\mu\nu}\dot{z}_j^{(2)\nu}(\tau_{\text{in},j})\\\nonumber&\approx m_j\eta_{\mu\nu}(z^{(2)\nu}_j(\tau_{\text{in},j}+\delta\tau)-b^\nu_j)/\delta\tau
\end{align}
 where we again have assumed that there is no gravitational field initially, $i.e.$ 
 \beq
 g_{\mu\nu}(z(\tau_{\text{in},j}))=\eta_{\mu\nu}~.
 \eeq
Using eq.~(\ref{delta(mu-P)}), we may then write the part of the expression in ~(\ref{dO-worldline-v2}) that we are focused on, in the time-discretized fashion
 \begin{align}\label{discretized-path}
     \int_{(t_\text{in},\vec{b}_j)}^{(t_\text{f},\vec{y}_j)}Dz_j^{(2)}|\tilde{\Phi}(\vec{\mathcal{P}}_{j,\text{in}}^{(2)})|^2&\propto\lim_{N\to\infty}\int\prod_{n=1}^N \{d^4z_j^{(2)}(\tau_{\text{in},j}+n\delta\tau)\}\delta^3(m_j\vec{u}_j-\vec{\mathcal{P}}^{(2)}_{j,\text{in}})
 \end{align}
 with $\delta\tau\sim (\tau_{\text{f},j}-\tau_{\text{in},j})/N$. Thus, when we replace $\vec{\mathcal{P}}^{(2)}_{j,\text{in}}$ in eq.~(\ref{discretized-path}) according to~(\ref{P-discrete}), we can regard the $\delta$-function in~(\ref{discretized-path}) as fixing the spatial part of the $n=1$ integration variable to be
\begin{equation}
    \vec{z}^{(2)}_j(\tau_{\text{in},j}+\delta\tau)=\vec{b}_j+\delta \tau\vec{u}_j
\end{equation}
Another way of phrasing this is that the presence of $|\tilde{\Phi}(\vec{\mathcal{P}}_{j,\text{in}}^{(2)})|^2$ in~(\ref{dO-worldline-v2}) ensures that we only integrate over paths for which
\begin{equation}
    \dot{\vec{z}}^{(2)}_j(\tau_{\text{in},j})=\vec{u}_j ~,
\end{equation}
as expected intuitively.

\subsubsection{Do we have enough boundary/initial conditions?}
 We have just uncovered how the initial conditions on $\vec{z}^{(2)}_j(\tau)$ are inherited directly from the KMOC wave functions in eq.~(\ref{KMOC-in-state}). The specific appearance of the copy "2" variables was in fact arbitrary and due to the order in which we performed the integrations. Still, are we guaranteed that this apparent asymmetry cures itself in the final answer? There are  other issues. For example, why is the time component of $z^{(2)\mu}_j(\tau_{\text{in},j}+\delta\tau)$ not restricted by any $\delta$-function constraint in~(\ref{discretized-path}) in a similar fashion as the spatial components?

These considerations raise the question  whether we have enough boundary conditions to specify a classical path. In other words, if we were to solve for the classical trajectory for both $z_j^{(1)\mu}$ and $z_j^{(2)\mu}$ by applying the principle of stationary phase to~(\ref{dO-worldline-v2}), what information should we supply as initial/boundary conditions for the equations of motion? The answer is the following: First the trajectory for $\vec{z}_j^{(2)}(\tau)$ is specified by the initial conditions from the KMOC wave functions 
 as just discussed, $i.e.$, $\vec{z}_j^{(2)}(\tau_{\text{in},j})=\vec{b}_j$ and $\dot{\vec{z}}_j^{(2)}(\tau_{\text{in},j})=\vec{u}_j$. We then note that this also determines a unique classical the spatial end-point $\vec{y}_{j,\text{cl}}=\vec{z}_{j,\text{cl}}^{(2)}(\tau_{\text{f},j})$, which according to eq.~(\ref{dO-worldline-v2}) is required to be the same for $\vec{z}_{j,\text{cl}}^{(1)}(\tau_{\text{f},j})$. So  $\vec{z}_j^{(1)}(\tau)$ is now uniquely determined by the boundary conditions $\vec{z}_j^{(1)}(\tau_{\text{in},j})=\vec{b}_j$ and $\vec{z}_j^{(1)}(\tau_{\text{f},j})=\vec{y}_{j,\text{cl}}$. This implies that the entire classical paths for the $(2)$ and $(1)$ variables are indeed the same because specifying the start-point and end-point determines a unique classical path. Of course, the fact that the trajectories are identical now also has the consequence that the initial classical velocity is the same $\dot{\vec{z}}_j^{(1)}(\tau_{\text{in},j})=\vec{u}_j$, even though this was never explicitly specified for the "1"-variables in~(\ref{dO-worldline-v2}). Finally, the classical zeroth components $z_j^{(i),0}(\tau)$ are uniquely specified by the boundary-conditions: $z_j^{(i),0}(\tau_{\text{in},j})=t_\text{in}$ and $z_j^{(i),0}(\tau_{\text{f},j})=t_\text{f}$ for $i=1,2$.\\
 
 In summary, we conclude that eq.~(\ref{dO-worldline-v2}) does contain all the information we need to calculate the classical paths for all variables  $z^{(i)\mu}_j(\tau)$. Specifically
\begin{eqnarray}
&& \vec{z}_j^{(2)}(\tau_{\text{in},j})=\vec{b}_j~,~
     \dot{\vec{z}}_j^{(2)}(\tau_{\text{in},j}) =\vec{u}_j~,~
     z_j^{(2),0}(\tau_{\text{in},j})=t_\text{in}~,~
     z_j^{(2),0}(\tau_{\text{f},j})=t_\text{f} \cr
&&  \vec{z}_j^{(1)}(\tau_{\text{in},j})=\vec{b}_j~,~
     \vec{z}_j^{(1)}(\tau_{\text{f},j})=\vec{z}_j^{(2)}(\tau_{\text{f},j})~,~
     z_j^{(1),0}(\tau_{\text{in},j})=t_\text{in}~,~
     z_j^{(1),0}(\tau_{\text{f},j})=t_\text{f}
\end{eqnarray}
which is equivalent to
\begin{eqnarray}\label{effective-conditions}
&&     \vec{z}_j^{(i)}(\tau_{\text{in},j})=\vec{b}_j~,~
     \dot{\vec{z}}_j^{(i)}(\tau_{\text{in},j}) =\vec{u}_j \cr
&&     z_j^{(i),0}(\tau_{\text{in},j})=t_\text{in}~,~
     z_j^{(i),0}(\tau_{\text{f},j})=t_\text{f}
 \end{eqnarray}
 \noindent for $i=1,2$.

 \subsubsection{Fixing the normalization}
Finally, we will fix the accumulation of overall constants in  $\mathcal{N}$ more explicitly. We first note that factors such as $E_j(\vec{\mathcal{P}}_{j,\text{f}}^{(2)})$  in 
the path integral~(\ref{dO-worldline-v2}) will have no influence on the classical solution for the trajectories since it is a smooth function of the worldline points and without oscillatory phases. This means that we can replace it by its classical value and move it outside the path integral where it can be absorbed into the overall constant. This kind of 
manipulation has already been done in other instances where we have discarded quantum effects from the path integral measure.

Following the discussion of boundary conditions above, let us introduce a simplified notation for the worldline path integral that expresses more clearly the fact that we are only integrating over paths consistent with the initial conditions from KMOC
\begin{align}
    \left .\int\limits_{\text{in-in}}{Dz_j^{(i)}}(...) \right|_{\substack{
    \dot{\vec{z}}_{j,\text{in}}^{(i)}=\vec{u}_j\\
    \vec{z}_{j,\text{in}}^{(i)}=\vec{b}_j}}\equiv \int d^3y_j\int^{z_j^{(i)}({\tau }_{\mathrm{f},j})=\left(t_{\mathrm{f}},\vec{y}_j\right)}_{z_j^{(i)}({\tau }_{\mathrm{in},j})=\left(t_{\mathrm{in}},\vec{b}_j\right)}{Dz_j^{(i)}}
 |\tilde{\Phi}_{j}(\vec{\mathcal{P}}_{j,\text{in}}^{(2)})|^2(...) .
\end{align}
With the help of this notation we can write~(\ref{dO-worldline-v2}) as
\begin{align}\label{dO-worldline-v3}
    \langle\Delta \hat{O}\rangle = \left.\mathcal{N}\int\limits_{\text{in-in}}{Dh^{(i)}}{Dz_j^{(i)}}\left(\int d\tau_1\frac{d}{d\tau_1}O_1\right)e^{\frac{i}{\hbar}(S[z_j^{\left(1\right)},h^{(1)}]-S[z_j^{\left(2\right)},h^{(2)}])}\right|_{\substack{
    \dot{\vec{z}}_{j,\text{in}}^{(i)}=\vec{u}_j\\
    \vec{z}_{j,\text{in}}^{(i)}=\vec{b}_j}}
\end{align}
We now fix $\mathcal{N}$ by unitarity,
\begin{align}
    1 = \langle \text{in}|\hat{S}^\dagger\hat{S}|\text{in}\rangle = \left.\mathcal{N} \int\limits_{
    \text{\text{in-in}
  }}{Dh^{(i)}Dz_j^{(i)}}e^{\frac{i}{\hbar}(S[z_j^{\left(1\right)},h^{(1)}]-S[z_j^{\left(2\right)},h^{(2)}])}\right|_{\substack{
    \dot{\vec{z}}_{j,\text{in}}^{(i)}=\vec{u}_j\\
    \vec{z}_{j,\text{in}}^{(i)}=\vec{b}_j}} ~,
\end{align}
which gives our final result

\begin{align}\label{dO-worldline-v4}
    {\langle\Delta \hat{O}\rangle = \left.\frac{\int\limits_{\text{in-in}}{Dh^{(i)}}{Dz_j^{(i)}}\left(\int d\tau_1\frac{d}{d\tau_1}O_1[\vec{\mathcal{P}}_1,\vec{z}_1]\right)e^{\frac{i}{\hbar}(S[z_j^{\left(1\right)},h^{(1)}]-S[z_j^{\left(2\right)},h^{(2)}])}}{\int\limits_{
    \text{\text{in-in}
  }}{Dh^{(i)}Dz_j^{(i)}}e^{\frac{i}{\hbar}(S[z_j^{\left(1\right)},h^{(1)}]-S[z_j^{\left(2\right)},h^{(2)}])}}\right|_{\substack{
    \dot{\vec{z}}_{j,\text{in}}^{(i)}=\vec{u}_j\\
    \vec{z}_{j,\text{in}}^{(i)}=\vec{b}_j}}}
\end{align}
Starting from the KMOC-expression for a general observable~(\ref{Delta-heuristic}), and restricting ourselves to the classical limit, we have thus explicitly mapped it to an {\em in-in} worldline expression with boundary conditions inherited from the KMOC wave functions. As stressed, we have demonstrated this equivalence only in the classical limit, given the numerous times terms of $\mathcal{O}(\hbar)$ have been neglected in the analysis.

\subsection{Momentum kick}

After this general prescription we now look specifically at the classical momentum kick which is related to the scattering angle. Picking one of the scalar particles, we therefore
choose the vector operator 
\begin{equation}\label{momentum}
    \hat{O}_1^{\mu}=\mathbb{P}^\mu_1=\sum_K \int d\mathrm{\Pi}_{p_1}d\mathrm{\Pi}_{p_2}\ket{p_1p_2;K}p_1^\mu\bra{p_1p_2;K} ~.
\end{equation}
Because  the on-shell $\delta$-functions were integrated out early on in the above analysis, the zeroth component of the momentum is identified as $p_1^0=E_1(p_1)=\sqrt{m_1^2+\vec{p}_1^2}$ everywhere. In order to see that this is equivalent to the worldline formalism, we need to identify $E_1(\vec{\mathcal{P}}_1)$ with 
\beq
\mathcal{P}^0_1~\equiv~ -\eta^{0\nu}\frac{\partial L_1}{\partial \dot{z}_1^{\nu}} ~.
\eeq
This identification follows from the fact that initial scalar particles are free particles plus the fact that the path integrals in~(\ref{dO-worldline-v4}) can be solved classically using the principle of stationary phase which replaces the variables with solutions to the classical equations of motion. We note that for a free particle the classical equations of motion together with the result in eq.~(\ref{v^2=1}) indeed imply that we can identify the zeroth component of the canonical momentum with $\mathcal{P}^0_1=E_1(\vec{\mathcal{P}}_1)$. Since only the end-points appear in~(\ref{dO-worldline-v4}) we have
\begin{equation}\label{momentum-v2}
    \int d\tau\frac{d}{d\tau}O_1[\vec{\mathcal{P}}_1,\vec{z}_1]^{\mu}=\int d\tau\frac{d}{d\tau}\mathcal{P}^\mu_1=-\eta^{\mu\nu}\int d\tau\frac{d}{d\tau}\frac{\partial L_1[z_1,\dot{z}_1;0]}{\partial \dot{z}_1^{\nu}}
\end{equation}
Inserting this into eq.~(\ref{dO-worldline-v4}) we then obtain
\begin{align}\label{final-result-momentum}
   \langle\Delta \mathbb{P}^\mu_1\rangle = -\eta^{\mu\nu}\left.\frac{\int\limits_{\text{in-in}}{Dh^{(i)}}{Dz_j^{(i)}}\int d\tau\frac{d}{d\tau}\frac{\partial L_1[z_1,\dot{z}_1;0]}{\partial \dot{z}_1^{\nu}}e^{\frac{i}{\hbar}(S[z_j^{\left(1\right)},h^{(1)}]-S[z_j^{\left(2\right)},h^{(2)}])}}{\int\limits_{
    \text{\text{in-in}
  }}{Dh^{(i)}Dz_j^{(i)}}e^{\frac{i}{\hbar}(S[z_j^{\left(1\right)},h^{(1)}]-S[z_j^{\left(2\right)},h^{(2)}])}}\right|_{\substack{
    \dot{\vec{z}}_{j,\text{in}}^{(i)}=\vec{u}_j\\
    \vec{z}_{j,\text{in}}^{(i)}=\vec{b}_j}}
\end{align}
which is the starting point for calculating the momentum kick from worldlines when dissipation is taken into account~\cite{Jakobsen:2022psy,Kalin:2022hph}. \\

For definiteness we will henceforth follow the effective field theory approach~\cite{Kalin:2022hph} for the practical calculations, although one could just as well have chosen the Worldline Quantum Field Theory formulation~\cite{Jakobsen:2022psy} to evaluate (\ref{final-result-momentum}). Let us first write 
(\ref{final-result-momentum}) as,
\begin{align}\label{final-result-momentum-eff}
    \langle\Delta \mathbb{P}^\mu_1\rangle = -\eta^{\mu\nu}\left.\frac{\int\limits_{\text{in-in}}{Dz_j^{(i)}}\left(\int d\tau\frac{d}{d\tau}\frac{\partial L_1[z_1,\dot{z}_1;0]}{\partial \dot{z}_1^{\nu}}\right)e^{\frac{i}{\hbar}S_{\text{eff}}[z_j^{\left(1\right)},z_j^{\left(2\right)}]}}{\int\limits_{
    \text{\text{in-in}
  }}{Dz_j^{(i)}}e^{\frac{i}{\hbar}S_{\text{eff}}[z_j^{\left(1\right)},z_j^{\left(2\right)}]}}\right|_{\substack{
    \dot{\vec{z}}_{j,\text{in}}^{(i)}=\vec{u}_j\\
    \vec{z}_{j,\text{in}}^{(i)}=\vec{b}_j}}
\end{align}
where the effective action $S_\text{eff}$ is defined by
\begin{align}
    e^{\frac{i}{\hbar}S_\text{eff}[z_j^{(1)},z_j^{(2)}]}&\equiv\int\limits_{\text{in-in}}{Dh^{(i)}}e^{\frac{i}{\hbar}(S[z_j^{\left(1\right)},h^{(1)}]-S[z_j^{\left(2\right)},h^{(2)}])}\\\nonumber
    &=\exp\left[\frac{i}{\hbar}\left[\int d\tau(L[z^{(1)},\dot{z}^{(1)};0]-L[z^{(2)},\dot{z}^{(2)};0])+S_\text{eff}^{(\text{int})}[z^{(1)},z^{(2)}]\right]\right]
\end{align}
We have separated out the two copies of the free part of the worldline Lagrangian $L[z^{(i)},\dot{z}^{(i)};0]=\sum_jL_j[z^{(i)}_j,\dot{z}^{(i)}_j;h_{\mu\nu}(z^{(i)}_j)=0]$. 
We now choose to evaluate the remaining path integral in~(\ref{final-result-momentum-eff}) by the principle of stationary phase, again ignoring quantum corrections. This replaces the worldline parameters $z_j^{(i)}(\tau)$ with their classical values, which we will denote $r_j(\tau)$. The expression thus becomes
\begin{equation}\label{clasical-impulse}
    \langle\Delta \mathbb{P}^\mu_1\rangle 
    = -\eta^{\mu\nu}\int d\tau\frac{d}{d\tau}\frac{\partial L_1[r_1,\dot{r}_1;0]}{\partial \dot{r}_1^{\nu}}
\end{equation}
The equations of motion providing $r_j(\tau)$ subject to the initial/boundary conditions stated above, can be written as
\beq
    0=\left.\frac{\delta S_\text{eff}[z^{(1)},z^{(2)}]}{\delta z_j^{(1)\mu}}\right|_{z_j^{(i)}=r_j}
    =\underbrace{\frac{\partial L_j[r_j,\dot{r}_j;0]}{\partial r_j^{\mu}}}_{=0\ (\text{cf. eq. (\ref{polyakov-lagrangian})})}-\frac{d}{d\tau}\frac{\partial L_j[r_j,\dot{r}_j;0]}{\partial \dot{r}_j^{\mu}}+\left.\frac{\delta S_\text{eff}^{(\text{int})}[z^{(1)},z^{(2)}]}{\delta z_j^{(1)\mu}}\right|_{z_j^{(i)}=r_j}
\eeq
Integrating this and using eq.~(\ref{clasical-impulse}), we obtain
\begin{equation}\label{dP_1-Seff}
    \langle\Delta \mathbb{P}^\mu_1\rangle=-\eta^{\mu\nu}\int d\tau\left.\frac{\delta S_\text{eff}^{(\text{int})}[z^{(1)},z^{(2)}]}{\delta z_1^{(1)\nu}}\right|_{z_j^{(i)}=r_j(\tau)}
\end{equation}
We now expand the (variation of the) effective action perturbatively. Using the notation defined in appendix~\ref{WEFT-Feynman-rules} (which is very similar to that of ref.~\cite{Dlapa:2021npj}), equation~(\ref{dP_1-Seff}) may be written as

\begin{align}\label{dP_diagrams}
   \langle\Delta \mathbb{P}^\mu_1\rangle=\int_{-\infty}^{\infty} d\tau \left\{\rule{0cm}{1.05cm}\right.&G\left[\rule{0cm}{1.05cm}\right. \vcenter{\hbox{\scalebox{0.7}{\begin{tikzpicture}
\begin{feynman}
\node [circle,draw=black, fill=white,inner sep=1pt,text width=3mm] (x);
\node [above=0.4cm of x ] (x1) {$\mu$};
\node [below=1.5cm of x,circle,draw=black, fill=black,inner sep=1pt,text width=3mm]  (y);
\node [below=0.3cm of y ] (y1);
\diagram*{
(y) -- [graviton,thick,opacity=0.7] (x),
(y) -- [fermion] (x),
};
\end{feynman}
\end{tikzpicture}}}}
+\vcenter{\hbox{\scalebox{0.7}{\begin{tikzpicture}
\begin{feynman}
\node [circle,draw=black, fill=white,inner sep=1pt,text width=3mm] (x);
\node [above=0.4cm of x ] (x1) {$\mu$};
\node [right=1.5cm of x,circle,draw=black, fill=black,inner sep=1pt,text width=3mm]  (y);
\node [above=0.3cm of y ] (y1);
\node [below=1.8cm ] (whale);
\diagram*{
(y) -- [graviton,thick,opacity=0.7,bend left] (x),
(y) -- [fermion,bend left] (x),
};
\end{feynman}
\end{tikzpicture}}}}\left.\rule{0cm}{1.05cm}\right]
+G^2\left[\rule{0cm}{1cm}\right.\vcenter{\hbox{\scalebox{0.7}{\begin{tikzpicture}
\begin{feynman}
\node [circle,draw=black, fill=white,inner sep=1pt,text width=3mm] (x);
\node [above=0.4cm of x ] (x1) {$\mu$};
\vertex [below=0.75cm of x]  (center);
\node [below=0.75cm of center ]  (whaleross);
\node [left=0.5cm of whaleross,circle,draw=black, fill=black,inner sep=1pt,text width=3mm]  (y-left);
\node [right=0.5cm of whaleross,circle,draw=black, fill=black,inner sep=1pt,text width=3mm]  (y-right);
\node [below=0.3cm of y-left ] (y1-left);
\node [below=0.3cm of y-right ] (y1-right);
\diagram*{
(center) -- [graviton,thick,opacity=0.7] (x),
(center) -- [fermion] (x),
(y-left) -- [graviton,thick,opacity=0.7] (center),
(y-left) -- [fermion] (center),
(y-right) -- [graviton,thick,opacity=0.7] (center),
(y-right) -- [fermion] (center),
};
\end{feynman}
\end{tikzpicture}}}}\\\nonumber&
+\vcenter{\hbox{\scalebox{0.7}{\begin{tikzpicture}
\begin{feynman}
\node [circle,draw=black, fill=black,inner sep=1pt,text width=3mm] (x);
\node [below=0.4cm of x ] (x1);
\vertex [above=0.75cm of x]  (center);
\node [above=0.75cm of center ]  (whaleross);
\node [left=0.5cm of whaleross,circle,draw=black, fill=white,inner sep=1pt,text width=3mm]  (y-left);
\node [right=0.5cm of whaleross,circle,draw=black, fill=black,inner sep=1pt,text width=3mm]  (y-right);
\node [above=0.3cm of y-left ] (y1-left) {$\mu$};
\node [above=0.3cm of y-right ] (y1-right);
\diagram*{
(x) -- [graviton,thick,opacity=0.7] (center),
(x) -- [fermion] (center),
(center) -- [graviton,thick,opacity=0.7] (y-left),
(center) -- [fermion] (y-left),
(y-right) -- [graviton,thick,opacity=0.7] (center),
(y-right) -- [fermion] (center),
};
\end{feynman}
\end{tikzpicture}}}}
+\vcenter{\hbox{\scalebox{0.7}{\begin{tikzpicture}
\begin{feynman}
\node [circle,draw=black, fill=black,inner sep=1pt,text width=3mm] (x);
\node [above=0.3cm of x ] (x1);
\vertex [below=0.7cm of x]  (center);
\node [left=0.8cm of x,circle,draw=black, fill=white,inner sep=1pt,text width=3mm]  (y-left);
\node [right=0.8cm of x,circle,draw=black, fill=black,inner sep=1pt,text width=3mm]  (y-right);
\node [above=0.3cm of y-left ] (y1-left) {$\mu$};
\node [above=0.3cm of y-right ] (y1-right);
\node [below=1.9 of x ] (invisible);
\diagram*{
(x) -- [graviton,thick,opacity=0.7] (center),
(x) -- [fermion] (center),
(center) -- [graviton,thick,opacity=0.7,bend left] (y-left),
(center) -- [fermion,bend left] (y-left),
(y-right) -- [graviton,thick,opacity=0.7,bend left] (center),
(y-right) -- [fermion,bend left] (center),
};
\end{feynman}
\end{tikzpicture}}}}\left.\rule{0cm}{1cm}\right]
+\mathcal{O}(G^3)\left.\rule{0cm}{1.05cm}\right\}
\end{align}
The vertex factors contain additional factors of $G$ since they depend on the classical trajectories. The classical trajectories can also be written in the same notation

\begin{align}\label{r1}
    r_1^\mu(\tau)=\frac{1}{m_1}\int_{-\infty}^{\tau} d\tau'\int_{-\infty}^{\tau'} d\tau'' \left\{\rule{0cm}{1.05cm}\right.&G\left[\rule{0cm}{1.05cm}\right. \vcenter{\hbox{\scalebox{0.7}{\begin{tikzpicture}
\begin{feynman}
\node [circle,draw=black, fill=white,inner sep=1pt,text width=3mm] (x);
\node [above=0.4cm of x ] (x1) {$\mu$};
\node [below=1.5cm of x,circle,draw=black, fill=black,inner sep=1pt,text width=3mm]  (y);
\node [below=0.3cm of y ] (y1);
\diagram*{
 (y) -- [graviton,thick,opacity=0.7] (x),
 (y) -- [fermion] (x),
};
\end{feynman}
\end{tikzpicture}}}}
+\vcenter{\hbox{\scalebox{0.7}{\begin{tikzpicture}
\begin{feynman}
\node [circle,draw=black, fill=white,inner sep=1pt,text width=3mm] (x);
\node [above=0.4cm of x ] (x1) {$\mu$};
\node [right=1.5cm of x,circle,draw=black, fill=black,inner sep=1pt,text width=3mm]  (y);
\node [above=0.3cm of y ] (y1);
\node [below=1.8cm ] (whale);
\diagram*{
 (y) -- [graviton,thick,opacity=0.7,bend left] (x),
 (y) -- [fermion,bend left] (x),
};
\end{feynman}
\end{tikzpicture}}}}\left.\rule{0cm}{1.05cm}\right]
+\mathcal{O}(G^2)\left.\rule{0cm}{1.05cm}\right\}
\end{align}
\begin{align}\label{r2}
    r_2^\mu(\tau)=\frac{1}{m_2}\int_{-\infty}^{\tau} d\tau'\int_{-\infty}^{\tau'} d\tau'' \left\{\rule{0cm}{1.05cm}\right.&G\left[\rule{0cm}{1.05cm}\right. \vcenter{\hbox{\scalebox{0.7}{\begin{tikzpicture}
\begin{feynman}
\node [circle,draw=black, fill=white,inner sep=1pt,text width=3mm] (x);
\node [below=0.4cm of x ] (x1) {$\mu$};
\node [above=1.5cm of x,circle,draw=black, fill=black,inner sep=1pt,text width=3mm]  (y);
\node [above=0.3cm of y ] (y1);
\diagram*{
 (y) -- [graviton,thick,opacity=0.7] (x),
 (y) -- [fermion] (x),
};
\end{feynman}
\end{tikzpicture}}}}
+\vcenter{\hbox{\scalebox{0.7}{\begin{tikzpicture}
\begin{feynman}
\node [circle,draw=black, fill=white,inner sep=1pt,text width=3mm] (x);
\node [below=0.4cm of x ] (x1) {$\mu$};
\node [right=1.5cm of x,circle,draw=black, fill=black,inner sep=1pt,text width=3mm]  (y);
\node [below=0.3cm of y ] (y1);
\node [above=1.8cm ] (whale);
\diagram*{
 (y) -- [graviton,thick,opacity=0.7,bend right] (x),
 (y) -- [fermion,bend right] (x),
};
\end{feynman}
\end{tikzpicture}}}}\left.\rule{0cm}{1.05cm}\right]
+\mathcal{O}(G^2)\left.\rule{0cm}{1.05cm}\right\}
\end{align}
These equations may be solved iteratively by writing
\begin{align}\label{r-expansion}
r_j^\mu(\tau)=\sum_{n=0}^\infty G^nr_{j,n}^\mu(\tau)
\end{align}
where the first order trajectory is determined by the initial conditions
\begin{align}
    r_{j,0}^\mu(\tau)=b_j^\mu+u_j^\mu\tau
\end{align}
We can then expand the vertices
\begin{align}\label{dP_diagrams2-expansion}
\vcenter{\hbox{\scalebox{0.9}{\begin{tikzpicture}
\begin{feynman}
\node (x);
\node [above=1cm of x,circle,draw=black, fill=white,inner sep=1pt,text width=3mm]  (y);
\node [below=0.2cm of x ] (x1);
\node [above=0.4cm of y ] (y1);
\diagram*{
 (y) -- [graviton,thick,opacity=0.7] (x),
 (y) -- [anti fermion] (x),
};
\end{feynman}
\end{tikzpicture}}}}=\sum_{n=0} G^n
\vcenter{\hbox{\scalebox{0.9}{\begin{tikzpicture}
\begin{feynman}
\node (x);
\node [above=1cm of x,circle,draw=black, fill=white,inner sep=1pt,text width=3mm]  (y) {$n$};
\node [below=0.2cm of x ] (x1);
\node [above=0.4cm of y ] (y1);
\diagram*{
 (y) -- [graviton,thick,opacity=0.7] (x),
 (y) -- [anti fermion] (x),
};
\end{feynman}
\end{tikzpicture}}}}
\end{align}
\begin{align}\label{dP_diagrams-expansion}
\vcenter{\hbox{\scalebox{0.9}{\begin{tikzpicture}
\begin{feynman}
\node (x);
\node [above=1cm of x,circle,draw=white, fill=black,inner sep=1pt,text width=3mm]  (y);
\node [below=0.2cm of x ] (x1);
\node [above=0.4cm of y ] (y1);
\diagram*{
 (y) -- [graviton,thick,opacity=0.7] (x),
 (y) -- [fermion] (x),
};
\end{feynman}
\end{tikzpicture}}}}=\sum_{n=0} G^n
\vcenter{\hbox{\scalebox{0.9}{\begin{tikzpicture}
\begin{feynman}
\node (x);
\node [above=1cm of x,circle,draw=white, fill=black,inner sep=1pt,text width=3mm]  (y) {\textcolor{white}{$n$}};
\node [below=0.2cm of x ] (x1);
\node [above=0.4cm of y ] (y1);
\diagram*{
 (y) -- [graviton,thick,opacity=0.7] (x),
 (y) -- [fermion] (x),
};
\end{feynman}
\end{tikzpicture}}}}
\end{align}

As an example the first order solutions may be written as (the self-force terms happen to be zero at this order)
\begin{align}\label{r11}
    r_{1,1}^\mu(\tau)=\frac{1}{m_1}\int_{-\infty}^{\tau} d\tau'\int_{-\infty}^{\tau'} d\tau''\vcenter{\hbox{\scalebox{0.7}{\begin{tikzpicture}
\begin{feynman}
\node [circle,draw=black, fill=white,inner sep=1pt,text width=3mm] (x) {$0$};
\node [above=0.4cm of x ] (x1) {$\mu$};
\node [below=1.5cm of x,circle,draw=black, fill=black,inner sep=1pt,text width=3mm]  (y) {\textcolor{white}{$0$}};
\node [below=0.3cm of y ] (y1);
\diagram*{
 (y) -- [graviton,thick,opacity=0.7] (x),
 (y) -- [fermion] (x),
};
\end{feynman}
\end{tikzpicture}}}}
\end{align}
\begin{align}\label{r21}
    r_{2,1}^\mu(\tau)=\frac{1}{m_2}\int_{-\infty}^{\tau} d\tau'\int_{-\infty}^{\tau'} d\tau'' \rule{0cm}{1.05cm} \vcenter{\hbox{\scalebox{0.7}{\begin{tikzpicture}
\begin{feynman}
\node [circle,draw=black, fill=white,inner sep=1pt,text width=3mm] (x) {$0$};
\node [below=0.4cm of x ] (x1) {$\mu$};
\node [above=1.5cm of x,circle,draw=black, fill=black,inner sep=1pt,text width=3mm]  (y) {\textcolor{white}{$0$}};
\node [above=0.3cm of y ] (y1);
\diagram*{
 (y) -- [graviton,thick,opacity=0.7] (x),
 (y) -- [fermion] (x),
};
\end{feynman}
\end{tikzpicture}}}}
\end{align}
and substituting the expansion~(\ref{r-expansion}) into the expression for the momentum change, gives the expansion of the momentum change to the given consistent order in $G$

\begin{align}\label{dP-iterated-diagrams}
   \langle\Delta \mathbb{P}^\mu_1\rangle=\int_{-\infty}^{\infty} d\tau \left\{\rule{0cm}{1.05cm}\right.&G\vcenter{\hbox{\scalebox{0.7}{\begin{tikzpicture}
\begin{feynman}
\node [circle,draw=black, fill=white,inner sep=1pt,text width=3mm] (x) {$0$};
\node [above=0.4cm of x ] (x1) {$\mu$};
\node [below=1.5cm of x,circle,draw=black, fill=black,inner sep=1pt,text width=3mm]  (y) {\textcolor{white}{$0$}};
\node [below=0.3cm of y ] (y1);
\diagram*{
(y) -- [graviton,thick,opacity=0.7] (x),
(y) -- [fermion] (x),
};
\end{feynman}
\end{tikzpicture}}}}
+G^2\left[\rule{0cm}{1cm}\right.
\vcenter{\hbox{\scalebox{0.7}{\begin{tikzpicture}
\begin{feynman}
\node [circle,draw=black, fill=white,inner sep=1pt,text width=3mm] (x) {$1$};
\node [above=0.4cm of x ] (x1) {$\mu$};
\node [below=1.5cm of x,circle,draw=black, fill=black,inner sep=1pt,text width=3mm]  (y) {\textcolor{white}{$0$}};
\node [below=0.3cm of y ] (y1);
\diagram*{
(y) -- [graviton,thick,opacity=0.7] (x),
(y) -- [fermion] (x),
};
\end{feynman}
\end{tikzpicture}}}}
+\vcenter{\hbox{\scalebox{0.7}{\begin{tikzpicture}
\begin{feynman}
\node [circle,draw=black, fill=white,inner sep=1pt,text width=3mm] (x) {$0$};
\node [above=0.4cm of x ] (x1) {$\mu$};
\node [below=1.5cm of x,circle,draw=black, fill=black,inner sep=1pt,text width=3mm]  (y){\textcolor{white}{$1$}};
\node [below=0.3cm of y ] (y1);
\diagram*{
(y) -- [graviton,thick,opacity=0.7] (x),
(y) -- [fermion] (x),
};
\end{feynman}
\end{tikzpicture}}}}
+\vcenter{\hbox{\scalebox{0.7}{\begin{tikzpicture}
\begin{feynman}
\node [circle,draw=black, fill=white,inner sep=1pt,text width=3mm] (x) {$0$};
\node [above=0.4cm of x ] (x1) {$\mu$};
\vertex [below=0.75cm of x]  (center);
\node [below=0.75cm of center ]  (whaleross);
\node [left=0.5cm of whaleross,circle,draw=black, fill=black,inner sep=1pt,text width=3mm]  (y-left){\textcolor{white}{$0$}};
\node [right=0.5cm of whaleross,circle,draw=black, fill=black,inner sep=1pt,text width=3mm]  (y-right){\textcolor{white}{$0$}};
\node [below=0.3cm of y-left ] (y1-left);
\node [below=0.3cm of y-right ] (y1-right);
\diagram*{
(center) -- [graviton,thick,opacity=0.7] (x),
(center) -- [fermion] (x),
(y-left) -- [graviton,thick,opacity=0.7] (center),
(y-left) -- [fermion] (center),
(y-right) -- [graviton,thick,opacity=0.7] (center),
(y-right) -- [fermion] (center),
};
\end{feynman}
\end{tikzpicture}}}}\\\nonumber&
+\vcenter{\hbox{\scalebox{0.7}{\begin{tikzpicture}
\begin{feynman}
\node [circle,draw=black, fill=black,inner sep=1pt,text width=3mm] (x){\textcolor{white}{$0$}};
\node [below=0.4cm of x ] (x1);
\vertex [above=0.75cm of x]  (center);
\node [above=0.75cm of center ]  (whaleross);
\node [left=0.5cm of whaleross,circle,draw=black, fill=white,inner sep=1pt,text width=3mm]  (y-left){$0$};
\node [right=0.5cm of whaleross,circle,draw=black, fill=black,inner sep=1pt,text width=3mm]  (y-right){\textcolor{white}{$0$}};
\node [above=0.3cm of y-left ] (y1-left);
\node [left=0.3cm of y1-left ] (x2) {$\mu$};
\node [above=0.3cm of y-right ] (y1-right);
\diagram*{
(x) -- [graviton,thick,opacity=0.7] (center),
(x) -- [fermion] (center),
(center) -- [graviton,thick,opacity=0.7] (y-left),
(center) -- [fermion] (y-left),
(y-right) -- [graviton,thick,opacity=0.7] (center),
(y-right) -- [fermion] (center),
};
\end{feynman}
\end{tikzpicture}}}}
\left.\rule{0cm}{1cm}\right]
+\mathcal{O}(G^3)\left.\rule{0cm}{1.05cm}\right\}
\end{align}

\section{The diagrammatic comparison to one-loop order}

Having just derived the equivalence between the KMOC and worldline formalisms it is of interest to see in detail how it
manifests itself in detailed calculations. Given that we had to make the formal rewritings at the level
of the $\hat{S}$-matrix itself, rather than specific representations of it, it should not be surprising that the equivalence
arises in a quite non-trivial manner order by order. We have already noted the cancellation of superclassical terms in 
the KMOC formalism, cancellations that are completely avoided in the wordline formalisms since their starting points
are manifestly classical to leading order in $\hbar$. 
As above, the particular observable we will be focusing on is the momentum transfer $\langle \Delta \mathbb{P}_1^\mu\rangle$ for one of the two matter particles. The KMOC expression can be written in terms of scattering amplitudes~\cite{Kosower:2018adc},

\beq\label{deltaP1}
    \langle\Delta \mathbb{P}^\mu_1\rangle =\frac{i}{\hbar} \langle\text{in}|[\mathbb{P}^\mu_1,\hat{T}]|\text{in}\rangle+
    \frac{1}{\hbar^2}\langle\text{in}|\hat{T}^{\dagger} [\mathbb{P}^\mu_1,\hat{T}]|\text{in}\rangle = I^\mu_{(1)}+I^\mu_{(2)}
\eeq
where the first term is given by
\begin{align}\label{I1}
    I^\mu_{(1)}=\int \frac{d^4q}{(2\pi)^2}&\left\{\rule{0cm}{1cm}\right.\theta(p_1^0+\hbar q)\theta(p_2^0-\hbar q) \delta(2p_1\cdot q+\hbar q^2)\\\nonumber&\times  \delta(2p_2\cdot q-\hbar q^2)e^{-ib\cdot q} i q^\mu\vcenter{\hbox{\scalebox{0.7}{\begin{tikzpicture}
\begin{feynman}
\vertex   (x);
\vertex [right=2cm of x]  (y);
\vertex [below=1.5cm of y]  (z);
\vertex [right=2cm of y]  (w);
\vertex [right=2cm of z]  (w1);
\vertex [left=2cm of z]  (x1);
\vertex [below=0.6cm of z] (invisible);
\vertex[below=0.5cm of y] (center);
\diagram*{
 (x) --  [momentum={[arrow shorten=0.12mm,arrow distance=2mm,xshift=-0.2cm]$p_1$}](y) ,
 (y) -- [momentum={[arrow shorten=0.12mm,arrow distance=2mm,xshift=0.5cm]$p_1+\hbar q$}](w),
 (x1) --  [momentum={[arrow shorten=0.12mm,arrow distance=2mm,xshift=-0.2cm]$p_2$}](z),
 (z) --  [momentum={[arrow shorten=0.12mm,arrow distance=2mm,xshift=0.5cm]$p_2-\hbar q$}](w1),
 (z) --[color=white] (invisible),
};
\end{feynman}
    \node[ellipse,
    draw,
    text = white,
    fill = gray,
    minimum width = 2.3cm, 
    minimum height = 0.8cm,rotate=90] (e)  at (2,-0.75) {\footnotesize All Diagrams} ;
\end{tikzpicture}}}}\left.\rule{0cm}{1cm}\right\}_{\substack{\mathrm{\scriptsize 
 expanded}\\ \text{to order } \hbar^0}}
\end{align}
and the second term in~(\ref{deltaP1}) is given by
\begin{align}\label{I2}
   I^\mu_{(2)}=&\sum_K\int \frac{d^4qd^4w_1d^4w_2}{(2\pi)^4}\left\{\rule{0cm}{1.4cm}\right.\theta(p_1^0+\hbar q^0)\theta(p_2^0-\hbar q^0) \delta(2p_1\cdot q+\hbar q^2)\\\nonumber&\times \delta(2p_2\cdot q-\hbar q^2)\theta(p_1^0+\hbar w_1^0)\theta(p_2^0+\hbar w_2^0) \delta(2p_1\cdot w_1+\hbar w^2_1)\delta(2p_2\cdot w_2+\hbar w^2_2)  \\\nonumber&\times e^{-ib\cdot q}\delta^4(w_1+w_2+\Sigma K) w^\mu_1 \hbar^{-1}\vcenter{\hbox{\scalebox{0.7}{\begin{tikzpicture}
\begin{feynman}
\vertex   (x);
\vertex [right=2cm of x]  (y);
\vertex [below=2cm of y,dot]  (z);
\vertex [right=2.7cm of y]  (w);
\vertex [right=2.7cm of z]  (w1);
\vertex [left=2cm of z]  (x1);
\vertex [below=0.6cm of z] (invisible);
\vertex [below=1cm of y] (center);
\vertex [right=2.3cm of center] (center-right) {...};
\vertex [above=0.3cm of center-right] (center-right-up);
\vertex [below=0.3cm of center-right] (center-right-down);
\diagram*{
(x) --  [momentum={[arrow shorten=0.12mm,arrow distance=2mm,xshift=-0.5cm]$p_1$}](y) ,
(y) -- [momentum={[arrow shorten=0.12mm,arrow distance=2mm]$p_1+\hbar w_1$}](w),
(x1) --  [momentum={[arrow shorten=0.12mm,arrow distance=2mm,xshift=-0.5cm]$p_2$}](z),
(z) --  [momentum={[arrow shorten=0.12mm,arrow distance=2mm]$p_2+\hbar w_2$}](w1),
(center) --[graviton,thick] (center-right-up),
(center) -- [draw=none,momentum={[arrow shorten=0.12mm,arrow distance=2mm,arrow style=gray]\footnotesize $\{K\}$}] (center-right-up),
(center) --[graviton,thick] (center-right),
(center) --[graviton,thick] (center-right-down),
};
\end{feynman}
   \node[ellipse,
   draw,
   text = white,
   fill = gray,
  minimum width = 3.4cm, 
   minimum height = 0.8cm,rotate=90] (e) at (2,-1) {\footnotesize All Diagrams};
\end{tikzpicture}}}}
\vcenter{\hbox{\scalebox{0.7}{\begin{tikzpicture}
\draw[gray, dashed] (0,1.5)--(0,-1.5);
\end{tikzpicture}}}}
\vcenter{\hbox{\scalebox{0.7}{\begin{tikzpicture}
\begin{feynman}
\vertex   (x);
\vertex [right=2.7cm of x]  (y);
\vertex [below=2cm of y,dot]  (z);
\vertex [right=2cm of y]  (w);
\vertex [right=2cm of z]  (w1);
\vertex [left=2.7cm of z]  (x1);
\vertex [below=0.6cm of z] (invisible);
\vertex[below=1cm of y] (center);
\vertex [left=2.4cm of center] (center-left) {...};
\vertex [above=0.3cm of center-left] (center-left-up);
\vertex [below=0.3cm of center-left] (center-left-down); 
\diagram*{
(x) --  [momentum={[arrow shorten=0.12mm,arrow distance=2mm]$p_1+\hbar w_1$}](y) ,
(y) -- [momentum={[arrow shorten=0.12mm,arrow distance=2mm,xshift=0.3cm]$p_1+\hbar q$}](w),
(x1) --  [momentum={[arrow shorten=0.12mm,arrow distance=2mm]$p_2+\hbar w_2$}](z),
(z) --  [momentum={[arrow shorten=0.12mm,arrow distance=2mm,xshift=0.3cm]$p_2-\hbar q$}](w1),
(center) -- [graviton,thick] (center-left-up),
(center-left-up) -- [draw=none,momentum={[arrow shorten=0.12mm,arrow distance=2mm,arrow style=gray]\footnotesize $\{K\}$}] (center),
(center) --[graviton,thick] (center-left),
(center) -- [graviton,thick] (center-left-down),
};
\end{feynman}
   \node[ellipse,
   draw,
   text = white,
   fill = gray,
   minimum width = 3.4cm, 
   minimum height = 0.8cm,rotate=90] (e) at (2.7,-1) {\footnotesize All Diagrams};
\end{tikzpicture}}}}\left.\rule{0cm}{1.5cm}\right\}_{\substack{\mathrm{\scriptsize 
expanded}\\ \text{to order } \hbar^0}}
\end{align}

\noindent where the amplitudes are defined as $T$-matrix elements,

\begin{equation}\label{amp}
\langle p_1^{'}p_2^{'};\{k\}|\hat{T}|p_1p_2;0 \rangle=(2\pi)^4\delta^4(p_1+p_2-p_1'-p_2'-\hbar \Sigma_jk_j)\vcenter{\hbox{\scalebox{0.7}{\begin{tikzpicture}
\begin{feynman}
\vertex   (x);
\vertex [right=2cm of x]  (y);
\vertex [below=2cm of y,dot]  (z);
\vertex [right=2.7cm of y]  (w);
\vertex [right=2.7cm of z]  (w1);
\vertex [left=2cm of z]  (x1);
\vertex [below=0.6cm of z] (invisible);
\vertex [below=1cm of y] (center);
\vertex [right=2.3cm of center] (center-right) {...};
\vertex [above=0.3cm of center-right] (center-right-up);
\vertex [below=0.3cm of center-right] (center-right-down);
\diagram*{
(x) --  [momentum={[arrow shorten=0.12mm,arrow distance=2mm,xshift=-0.5cm]$p_1$}](y) ,
(y) -- [momentum={[arrow shorten=0.12mm,arrow distance=2mm]$p_1'$}](w),
(x1) --  [momentum={[arrow shorten=0.12mm,arrow distance=2mm,xshift=-0.5cm]$p_2$}](z),
(z) --  [momentum={[arrow shorten=0.12mm,arrow distance=2mm]$p_2'$}](w1),
(center) --[graviton,thick] (center-right-up),
(center) -- [draw=none,momentum={[arrow shorten=0.12mm,arrow distance=2mm,arrow style=gray]\footnotesize $\{\hbar k\}$}] (center-right-up),
(center) --[graviton,thick] (center-right),
(center) --[graviton,thick] (center-right-down),
};
\end{feynman}
  \node[ellipse,
   draw,
  text = white,
   fill = gray,
  minimum width = 3.4cm, 
   minimum height = 0.8cm,rotate=90] (e) at (2,-1) {\footnotesize All Diagrams};
\end{tikzpicture}}}}
\end{equation}

Here the scalar momenta are $p_1 = m_1u_1, p_2 = m_2u_2$ and the impact parameter is $b \equiv b_1-b_2$. The symbolic sum over $K$ is again a short-hand for the combined sum and integrations over any number of graviton degrees of freedom. In general, the extraction of the classical contribution requires us to expand both the amplitudes and all other factors contained in the expressions~(\ref{I1}) and~(\ref{I2}). As is well-known, there will also be superclassical terms, which only cancel once all contributions are considered.

For illustration, we consider here the Einstein-Hilbert action fixed to harmonic gauge and coupled to the two scalars in standard fashion,
\begin{align}\label{full-action}
S=\int d^4x &\left[\frac{\sqrt{-g}R}{16\pi G}+\left(\partial_\gamma h^{\gamma\mu}-\frac{1}{2}\partial^\mu h\right)^2\right.\\\nonumber&\left.+\sum_{j=1}^2\frac{1}{2}\sqrt{-g}\left(g^{\mu\nu}\partial_\mu\phi_j\partial_\nu\phi_j-\frac{m_j^2}{\hbar^2}\phi_j^2\right)\right],
\end{align}
where $g_{\mu\nu}=\eta_{\mu\nu}+\sqrt{32\pi G}h_{\mu\nu}$, $R$ is the Ricci scalar, and $\eta_{\mu\nu}$ is the Minkowski metric with $(+---)$ signature. A brief note on our 
$\hbar$-power-counting: Feynman rules with the present conventions assign to each graviton-scalar vertex a factor of $1/\hbar^3$, each scalar propagator comes with a factor of $\hbar^3$, each graviton three-vertex comes with a factor of $1/\hbar$, and each graviton propagator carries a factor of $\hbar$ (if external gravitons were considered, they would each carry a factor of $\sqrt{\hbar}$). Every diagram is multiplied by an overall factor of $\hbar^5$.

\subsection{Leading order comparison}
Interestingly, already at tree level the Feynman diagram approach leads to the result in a way that is quite distinct from
the worldline perspective. The only contribution to the KMOC expression at order $G$ is
\begin{align}\label{dP-1PM-KMOC}
    \langle\Delta \mathbb{P}^\mu_1\rangle_{\text{\tiny 1PM}}&=\int \frac{d^4q}{(2\pi)^2}\frac{\delta(p_1\cdot q)\delta(p_2\cdot q)}{4}e^{-ib\cdot q} i q^\mu \left\{\vcenter{\hbox{\scalebox{0.8}{\begin{tikzpicture}
\begin{feynman}
\vertex   (x);
\vertex [right=1.5cm of x]  (y);
\vertex [below=1.5cm of y]  (z);
\vertex [right=1.5cm of y]  (w);
\vertex [right=1.5cm of z]  (w1);
\vertex [left=1.5cm of z]  (x1);
\vertex [below=0.6cm of z] (invisible);
\vertex[below=0.5cm of y] (center);
\diagram*{
 (x) --  [momentum={[arrow shorten=0.12mm,arrow distance=2mm,xshift=-0.2cm]$p_1$}](y) ,
 (y) -- [momentum={[arrow shorten=0.12mm,arrow distance=2mm,xshift=0.5cm]$p_1+\hbar q$}](w),
 (x1) --  [momentum={[arrow shorten=0.12mm,arrow distance=2mm,xshift=-0.2cm]$p_2$}](z),
 (z) --  [momentum={[arrow shorten=0.12mm,arrow distance=2mm,xshift=0.5cm]$p_2-\hbar q$}](w1),
 (z) --[color=white] (invisible),
 (z) --[graviton] (y),
};
\end{feynman}
\end{tikzpicture}}}}\right\}_{\mathcal{O}(\hbar^0)}\\\nonumber
 &=(8\pi Gm_1m_2)\left[\frac{1}{2}(2\gamma^2-1)\right]\int\frac{d^4q}{(2\pi)^2}
\delta(u_1\cdot q)\delta(u_2\cdot q)e^{ib\cdot q}\frac{(-iq^{\mu})}{q^2} \\\nonumber
&= - (4\pi Gm_1m_2)\left[\frac{1}{\beta\gamma}(2\gamma^2-1)\right]\delta^{\mu}_{j}\int\frac{d^2q_{\perp}}{(2\pi)^2}
e^{ib_{\perp}\cdot q}\frac{iq_{\perp}^j}{q_{\perp}^2} \\\nonumber
&= -2Gm_1m_2\frac{2\gamma^2-1}{\sqrt{\gamma^2 -1}}\frac{\delta^{\mu}_{j}b_{\perp}^j}{b_{\perp}^2}
\end{align}
after inserting $\beta\gamma = \sqrt{\gamma^2-1}$, where $\gamma\equiv u_1\cdot u_2 $. In the third line we evaluated the expression in the rest frame of particle 1. The two $\delta$-function constraints in the above integral arise
from the on-shell Lorentz-invariant phase space integrations in the KMOC-formula, ignoring quantum terms. To this leading
1PM order this is intuitively understandable and should correspond to treating the two scalars as classical objects moving
on-shell. We can already anticipate that this interpretation becomes more subtle once we start to include loops in
the Feynman diagram expansion.

Let us compare the tree-level computation with the corresponding leading-order worldline calculation
\begin{align}\label{dP-1PM-wl}
    \int_{-\infty}^{\infty} d\tau G
\vcenter{\hbox{\scalebox{0.8}{\begin{tikzpicture}
\begin{feynman}
\node [circle,draw=black, fill=white,inner sep=1pt,text width=3mm] (x) {$0$};
\node [above=0.4cm of x ] (x1) {$\mu$};
\node [below=1.5cm of x,circle,draw=black, fill=black,inner sep=1pt,text width=3mm]  (y) {\textcolor{white}{$0$}};
\node [below=0.3cm of y ] (y1);
\diagram*{
 (y) -- [graviton,thick,opacity=0.7] (x),
 (y) -- [fermion] (x),
};
\end{feynman}
\end{tikzpicture}}}}
&=8\pi G m_1m_2\left[\frac{1}{2}(2\gamma^2-1)\right] \int \frac{d^4k}{(2\pi)^4}\int d\tau d\tau' e^{-iu_1\cdot k\tau}e^{iu_2\cdot k\tau'}e^{-ib\cdot k}\frac{ik^\mu}{k^2}\\\nonumber
&=8\pi G m_1m_2\left[\frac{1}{2}(2\gamma^2-1)\right] \int \frac{d^4k}{(2\pi)^2}\delta(u_1\cdot k)\delta(u_2\cdot k)e^{ib\cdot k}\frac{-ik^\mu}{k^2}\\\nonumber
 &=\langle\Delta \mathbb{P}^\mu_1\rangle_{\text{\tiny 1PM}}
\end{align}
The first observation is that the on-shell
$\delta$-function constraints arise in a completely different manner, now from the Fourier transform of the (leading-order)
straight-line trajectories. Apart from this, the Feynman rule for the worldline-worldline-graviton vertex matches closely
that of the second quantized field theory to this order. The different $i\epsilon$-prescriptions clearly play no role at this order.

\subsection{Next-to-leading order comparison}
Already at one-loop order the correspondence gets more intricate. We first draw the different Feynman diagrams that contribute
to this order in the KMOC-formalism. The calculation is quite standard and can be found in many places in the literature but, we find it illuminating to write out all
terms in order to explore the relationship to the corresponding worldline calculation. We list the Feynman diagrams here
\begin{align}\label{dP-2PM-KMOC-V}
    \langle\Delta \mathbb{P}^\mu_1\rangle_{\text{\tiny 2PM,$V$}}&=\int \frac{d^4q}{(2\pi)^2}\frac{\delta(p_1\cdot q)\delta(p_2\cdot q)}{4}e^{-ib\cdot q} i q^\mu \left\{\vcenter{\hbox{\scalebox{0.8}{\begin{tikzpicture}
\begin{feynman}
\vertex   (x);
\vertex [right=1.5cm of x]  (u);
\vertex [right=0.5cm of u]  (y);
\vertex [right=0.5cm of y]  (u1);
\vertex [below=1.5cm of y]  (z);
\vertex [right=1.5cm of u1]  (w);
\vertex [right=2cm of z]  (w1);
\vertex [left=2cm of z]  (x1);
\vertex [below=0.6cm of z] (invisible);
\vertex[below=0.5cm of y] (center);
\diagram*{
 (x) --  [momentum={[arrow shorten=0.12mm,arrow distance=2mm,xshift=-0.2cm]$p_1$}](u) ,
 (u) --  (u1) ,
 (u1) -- [momentum={[arrow shorten=0.12mm,arrow distance=2mm,xshift=0.5cm]$p_1+\hbar q$}](w),
 (x1) --  [momentum={[arrow shorten=0.12mm,arrow distance=2mm,xshift=-0.2cm]$p_2$}](z),
 (z) --  [momentum={[arrow shorten=0.12mm,arrow distance=2mm,xshift=0.5cm]$p_2-\hbar q$}](w1),
 (z) --[color=white] (invisible),
 (z) --[graviton] (u),
 (z) --[graviton] (u1),
};
\end{feynman}
\end{tikzpicture}}}}\right\}_{\mathcal{O}(\hbar^0)}\\\nonumber
\end{align}
\begin{align}\label{dP-2PM-KMOC-Lam}
    \langle\Delta \mathbb{P}^\mu_1\rangle_{\text{\tiny 2PM,$\Lambda$}}&=\int \frac{d^4q}{(2\pi)^2}\frac{\delta(p_1\cdot q)\delta(p_2\cdot q)}{4}e^{-ib\cdot q} i q^\mu \left\{\vcenter{\hbox{\scalebox{0.8}{\begin{tikzpicture}
\begin{feynman}
\vertex   (x);
\vertex [right=1.5cm of x]  (u);
\vertex [right=0.5cm of u]  (y);
\vertex [right=0.5cm of y]  (u1);
\vertex [above=1.5cm of y]  (z);
\vertex [right=1.5cm of u1]  (w);
\vertex [right=2cm of z]  (w1);
\vertex [left=2cm of z]  (x1);
\vertex [below=0.6cm of y] (invisible);
\vertex[above=0.5cm of y] (center);
\diagram*{
 (x) --  [momentum={[arrow shorten=0.12mm,arrow distance=2mm,xshift=-0.2cm]$p_2$}](u) ,
 (u) --  (u1) ,
 (u1) -- [momentum={[arrow shorten=0.12mm,arrow distance=2mm,xshift=0.5cm]$p_2-\hbar q$}](w),
 (x1) --  [momentum={[arrow shorten=0.12mm,arrow distance=2mm,xshift=-0.2cm]$p_1$}](z),
 (z) --  [momentum={[arrow shorten=0.12mm,arrow distance=2mm,xshift=0.5cm]$p_1+\hbar q$}](w1),
 (z) --[color=white] (invisible),
 (z) --[graviton] (u),
 (z) --[graviton] (u1),
};
\end{feynman}
\end{tikzpicture}}}}\right\}_{\mathcal{O}(\hbar^0)}\\\nonumber
\end{align}

\begin{align}\label{dP-2PM-KMOC-Y}
    \langle\Delta \mathbb{P}^\mu_1\rangle_{\text{\tiny 2PM,$Y$}}&=\int \frac{d^4q}{(2\pi)^2}\frac{\delta(p_1\cdot q)\delta(p_2\cdot q)}{4}e^{-ib\cdot q} i q^\mu \left\{\vcenter{\hbox{\scalebox{0.8}{\begin{tikzpicture}
\begin{feynman}
\vertex   (x);
\vertex [right=1.5cm of x]  (u);
\vertex [right=0.5cm of u]  (y);
\vertex [right=0.5cm of y]  (u1);
\vertex [below=1.5cm of y]  (z);
\vertex [right=1.5cm of u1]  (w);
\vertex [right=2cm of z]  (w1);
\vertex [left=2cm of z]  (x1);
\vertex [below=0.6cm of z] (invisible);
\vertex[below=0.75cm of y] (center);
\diagram*{
 (x) --  [momentum={[arrow shorten=0.12mm,arrow distance=2mm,xshift=-0.2cm]$p_1$}](u) ,
 (u) --  (u1) ,
 (u1) -- [momentum={[arrow shorten=0.12mm,arrow distance=2mm,xshift=0.5cm]$p_1+\hbar q$}](w),
 (x1) --  [momentum={[arrow shorten=0.12mm,arrow distance=2mm,xshift=-0.2cm]$p_2$}](z),
 (z) --  [momentum={[arrow shorten=0.12mm,arrow distance=2mm,xshift=0.5cm]$p_2-\hbar q$}](w1),
 (z) --[color=white] (invisible),
 (center) --[graviton] (u),
 (center) --[graviton] (u1),
 (z) --[graviton] (center),
};
\end{feynman}
\end{tikzpicture}}}}\right\}_{\mathcal{O}(\hbar^0)}\\\nonumber
\end{align}

\begin{align}\label{dP-2PM-KMOC-Y'}
    \langle\Delta \mathbb{P}^\mu_1\rangle_{\text{\tiny 2PM,$Y$'}}&=\int \frac{d^4q}{(2\pi)^2}\frac{\delta(p_1\cdot q)\delta(p_2\cdot q)}{4}e^{-ib\cdot q} i q^\mu \left\{\vcenter{\hbox{\scalebox{0.8}{\begin{tikzpicture}
\begin{feynman}
\vertex   (x);
\vertex [right=1.5cm of x]  (u);
\vertex [right=0.5cm of u]  (y);
\vertex [right=0.5cm of y]  (u1);
\vertex [above=1.5cm of y]  (z);
\vertex [right=1.5cm of u1]  (w);
\vertex [right=2cm of z]  (w1);
\vertex [left=2cm of z]  (x1);
\vertex [below=0.6cm of y] (invisible);
\vertex[above=0.75cm of y] (center);
\diagram*{
 (x) --  [momentum={[arrow shorten=0.12mm,arrow distance=2mm,xshift=-0.2cm]$p_2$}](u) ,
 (u) --  (u1) ,
 (u1) -- [momentum={[arrow shorten=0.12mm,arrow distance=2mm,xshift=0.5cm]$p_2-\hbar q$}](w),
 (x1) --  [momentum={[arrow shorten=0.12mm,arrow distance=2mm,xshift=-0.2cm]$p_1$}](z),
 (z) --  [momentum={[arrow shorten=0.12mm,arrow distance=2mm,xshift=0.5cm]$p_1+\hbar q$}](w1),
 (z) --[color=white] (invisible),
 (center) --[graviton] (u),
 (center) --[graviton] (u1),
 (center) --[graviton] (z),
};
\end{feynman}
\end{tikzpicture}}}}\right\}_{\mathcal{O}(\hbar^0)}\\\nonumber
\end{align}

\begin{align}\label{dP-2PM-KMOC-B}
    \langle\Delta \mathbb{P}^\mu_1\rangle_{\text{\tiny 2PM,$B.1$}}&=\int \frac{d^4q}{(2\pi)^2}\frac{\delta'(p_1\cdot q)\delta(p_2\cdot q)}{8}q^2e^{-ib\cdot q} i q^\mu\\\nonumber &\times \hbar \left\{\vcenter{\hbox{\scalebox{0.8}{\begin{tikzpicture}
\begin{feynman}
\vertex   (x);
\vertex [right=1.5cm of x]  (u);
\vertex [right=0.5cm of u]  (y);
\vertex [right=0.5cm of y]  (u1);
\vertex [above=1.5cm of y]  (z);
\vertex [right=1.5cm of u1]  (w);
\vertex [right=0.5cm of z]  (r1);
\vertex [left=0.5cm of z]  (l1);
\vertex [right=2cm of z]  (w1);
\vertex [left=2cm of z]  (x1);
\vertex [below=0.6cm of y] (invisible);
\vertex[above=0.75cm of y] (center);
\diagram*{
 (x) --  [momentum={[arrow shorten=0.12mm,arrow distance=2mm,xshift=-0.2cm]$p_2$}](u) ,
 (u) --  (u1) ,
 (u1) -- [momentum={[arrow shorten=0.12mm,arrow distance=2mm,xshift=0.5cm]$p_2-\hbar q$}](w),
 (x1) --  [momentum={[arrow shorten=0.12mm,arrow distance=2mm,xshift=-0.2cm]$p_1$}](z),
 (z) --  [momentum={[arrow shorten=0.12mm,arrow distance=2mm,xshift=0.5cm]$p_1+\hbar q$}](w1),
 (z) --[color=white] (invisible),
 (u1) --[graviton] (r1),
 (u) --[graviton] (l1),
};
\end{feynman}
\end{tikzpicture}}}}+\vcenter{\hbox{\scalebox{0.8}{\begin{tikzpicture}
\begin{feynman}
\vertex   (x);
\vertex [right=1.5cm of x]  (u);
\vertex [right=0.5cm of u]  (y);
\vertex [right=0.5cm of y]  (u1);
\vertex [above=1.5cm of y]  (z);
\vertex [right=1.5cm of u1]  (w);
\vertex [right=0.5cm of z]  (r1);
\vertex [left=0.5cm of z]  (l1);
\vertex [right=2cm of z]  (w1);
\vertex [left=2cm of z]  (x1);
\vertex [below=0.6cm of y] (invisible);
\vertex[above=0.75cm of y] (center);
\diagram*{
 (x) --  [momentum={[arrow shorten=0.12mm,arrow distance=2mm,xshift=-0.2cm]$p_2$}](u) ,
 (u) --  (u1) ,
 (u1) -- [momentum={[arrow shorten=0.12mm,arrow distance=2mm,xshift=0.5cm]$p_2-\hbar q$}](w),
 (x1) --  [momentum={[arrow shorten=0.12mm,arrow distance=2mm,xshift=-0.2cm]$p_1$}](z),
 (z) --  [momentum={[arrow shorten=0.12mm,arrow distance=2mm,xshift=0.5cm]$p_1+\hbar q$}](w1),
 (z) --[color=white] (invisible),
 (u1) --[graviton] (l1),
 (u) --[graviton] (r1),
};
\end{feynman}
\end{tikzpicture}}}}\right\}_{\mathcal{O}(\hbar^{-1})}\\\nonumber
\end{align}

\begin{align}\label{dP-2PM-KMOC-B2}
    \langle\Delta \mathbb{P}^\mu_1\rangle_{\text{\tiny 2PM,$B.2$}}&=-\int \frac{d^4q}{(2\pi)^2}\frac{\delta(p_1\cdot q)\delta'(p_2\cdot q)}{8}q^2e^{-ib\cdot q} i q^\mu\\\nonumber &\times \hbar \left\{\vcenter{\hbox{\scalebox{0.8}{\begin{tikzpicture}
\begin{feynman}
\vertex   (x);
\vertex [right=1.5cm of x]  (u);
\vertex [right=0.5cm of u]  (y);
\vertex [right=0.5cm of y]  (u1);
\vertex [above=1.5cm of y]  (z);
\vertex [right=1.5cm of u1]  (w);
\vertex [right=0.5cm of z]  (r1);
\vertex [left=0.5cm of z]  (l1);
\vertex [right=2cm of z]  (w1);
\vertex [left=2cm of z]  (x1);
\vertex [below=0.6cm of y] (invisible);
\vertex[above=0.75cm of y] (center);
\diagram*{
 (x) --  [momentum={[arrow shorten=0.12mm,arrow distance=2mm,xshift=-0.2cm]$p_2$}](u) ,
 (u) --  (u1) ,
 (u1) -- [momentum={[arrow shorten=0.12mm,arrow distance=2mm,xshift=0.5cm]$p_2-\hbar q$}](w),
 (x1) --  [momentum={[arrow shorten=0.12mm,arrow distance=2mm,xshift=-0.2cm]$p_1$}](z),
 (z) --  [momentum={[arrow shorten=0.12mm,arrow distance=2mm,xshift=0.5cm]$p_1+\hbar q$}](w1),
 (z) --[color=white] (invisible),
 (u1) --[graviton] (r1),
 (u) --[graviton] (l1),
};
\end{feynman}
\end{tikzpicture}}}}+\vcenter{\hbox{\scalebox{0.8}{\begin{tikzpicture}
\begin{feynman}
\vertex   (x);
\vertex [right=1.5cm of x]  (u);
\vertex [right=0.5cm of u]  (y);
\vertex [right=0.5cm of y]  (u1);
\vertex [above=1.5cm of y]  (z);
\vertex [right=1.5cm of u1]  (w);
\vertex [right=0.5cm of z]  (r1);
\vertex [left=0.5cm of z]  (l1);
\vertex [right=2cm of z]  (w1);
\vertex [left=2cm of z]  (x1);
\vertex [below=0.6cm of y] (invisible);
\vertex[above=0.75cm of y] (center);
\diagram*{
 (x) --  [momentum={[arrow shorten=0.12mm,arrow distance=2mm,xshift=-0.2cm]$p_2$}](u) ,
 (u) --  (u1) ,
 (u1) -- [momentum={[arrow shorten=0.12mm,arrow distance=2mm,xshift=0.5cm]$p_2-\hbar q$}](w),
 (x1) --  [momentum={[arrow shorten=0.12mm,arrow distance=2mm,xshift=-0.2cm]$p_1$}](z),
 (z) --  [momentum={[arrow shorten=0.12mm,arrow distance=2mm,xshift=0.5cm]$p_1+\hbar q$}](w1),
 (z) --[color=white] (invisible),
 (u1) --[graviton] (l1),
 (u) --[graviton] (r1),
};
\end{feynman}
\end{tikzpicture}}}}\right\}_{\mathcal{O}(\hbar^{-1})}\\\nonumber
\end{align}
We note that the only non-vanishing classical contribution from the box plus crossed box diagrams in equations~(\ref{dP-2PM-KMOC-B}) and~(\ref{dP-2PM-KMOC-B2}) arise from expanding the on-shell delta functions in eq.~(\ref{I1}) to first order in $\hbar$. When multiplied by the superclassical part of the amplitude of order ($\hbar^{-1}$) this provides
at term of order ($\hbar^{0}$). All other contributions than those listed above vanish. However, and as is well-known, there are also non-vanishing superclassical terms produced by these diagrams,

\begin{align}\label{dP-2PM-KMOC-super}
    \langle\Delta \mathbb{P}^\mu_1\rangle_{\text{\tiny 2PM,$B,\hbar^{-1}$}}&=\int \frac{d^4q}{(2\pi)^2}\frac{\delta(p_1\cdot q)\delta(p_2\cdot q)}{4}e^{-ib\cdot q} i q^\mu\\\nonumber &\times \left\{\vcenter{\hbox{\scalebox{0.8}{\begin{tikzpicture}
\begin{feynman}
\vertex   (x);
\vertex [right=1.5cm of x]  (u);
\vertex [right=0.5cm of u]  (y);
\vertex [right=0.5cm of y]  (u1);
\vertex [above=1.5cm of y]  (z);
\vertex [right=1.5cm of u1]  (w);
\vertex [right=0.5cm of z]  (r1);
\vertex [left=0.5cm of z]  (l1);
\vertex [right=2cm of z]  (w1);
\vertex [left=2cm of z]  (x1);
\vertex [below=0.6cm of y] (invisible);
\vertex[above=0.75cm of y] (center);
\diagram*{
 (x) --  [momentum={[arrow shorten=0.12mm,arrow distance=2mm,xshift=-0.2cm]$p_2$}](u) ,
 (u) --  (u1) ,
 (u1) -- [momentum={[arrow shorten=0.12mm,arrow distance=2mm,xshift=0.5cm]$p_2-\hbar q$}](w),
 (x1) --  [momentum={[arrow shorten=0.12mm,arrow distance=2mm,xshift=-0.2cm]$p_1$}](z),
 (z) --  [momentum={[arrow shorten=0.12mm,arrow distance=2mm,xshift=0.5cm]$p_1+\hbar q$}](w1),
 (z) --[color=white] (invisible),
 (u1) --[graviton] (r1),
 (u) --[graviton] (l1),
};
\end{feynman}
\end{tikzpicture}}}}+\vcenter{\hbox{\scalebox{0.8}{\begin{tikzpicture}
\begin{feynman}
\vertex   (x);
\vertex [right=1.5cm of x]  (u);
\vertex [right=0.5cm of u]  (y);
\vertex [right=0.5cm of y]  (u1);
\vertex [above=1.5cm of y]  (z);
\vertex [right=1.5cm of u1]  (w);
\vertex [right=0.5cm of z]  (r1);
\vertex [left=0.5cm of z]  (l1);
\vertex [right=2cm of z]  (w1);
\vertex [left=2cm of z]  (x1);
\vertex [below=0.6cm of y] (invisible);
\vertex[above=0.75cm of y] (center);
\diagram*{
 (x) --  [momentum={[arrow shorten=0.12mm,arrow distance=2mm,xshift=-0.2cm]$p_2$}](u) ,
 (u) --  (u1) ,
 (u1) -- [momentum={[arrow shorten=0.12mm,arrow distance=2mm,xshift=0.5cm]$p_2-\hbar q$}](w),
 (x1) --  [momentum={[arrow shorten=0.12mm,arrow distance=2mm,xshift=-0.2cm]$p_1$}](z),
 (z) --  [momentum={[arrow shorten=0.12mm,arrow distance=2mm,xshift=0.5cm]$p_1+\hbar q$}](w1),
 (z) --[color=white] (invisible),
 (u1) --[graviton] (l1),
 (u) --[graviton] (r1),
};
\end{feynman}
\end{tikzpicture}}}}\right\}_{\mathcal{O}(\hbar^{-1})}\\\nonumber
\end{align}
These should be cancelled by the subtractions due to the second term in eq.~(\ref{deltaP1}),

\begin{align}\label{dP-KMOC-A}
    \langle\Delta \mathbb{P}^\mu_1\rangle_{\text{\tiny 2PM,$A,\hbar^{-1}$}}&=\int \frac{d^4qd^4w}{(2\pi)^4}\frac{\delta(p_1\cdot q) \delta(p_2\cdot q) \delta(p_1\cdot w)\delta(p_2\cdot w)}{16} e^{-ib\cdot q} w^\mu\\\nonumber&\times \frac{1}{\hbar}\left\{\vcenter{\hbox{\scalebox{0.7}{\begin{tikzpicture}
\begin{feynman}
\vertex   (x);
\vertex [right=2cm of x]  (y);
\vertex [below=1.5cm of y]  (z);
\vertex [right=2.7cm of y]  (w);
\vertex [right=2.7cm of z]  (w1);
\vertex [left=2cm of z]  (x1);
\vertex [below=0.6cm of z] (invisible);
\vertex [below=0.75cm of y] (center);
\vertex [right=2.5cm of center] (center-right) {$\times$};
\vertex [above=0.3cm of center-right] (center-right-up);
\vertex [below=0.3cm of center-right] (center-right-down);
\vertex [below=0.6cm of z] (invisible);
\diagram*{
 (x) --  [momentum={[arrow shorten=0.12mm,arrow distance=2mm,xshift=-0.5cm]$p_1$}](y) ,
 (y) -- [momentum={[arrow shorten=0.12mm,arrow distance=2mm]$p_1+\hbar w$}](w),
 (x1) --  [momentum={[arrow shorten=0.12mm,arrow distance=2mm,xshift=-0.5cm]$p_2$}](z),
 (z) --  [momentum={[arrow shorten=0.12mm,arrow distance=2mm]$p_2-\hbar w$}](w1),
 (z) --[graviton](y),
 (z) --[color=white] (invisible),
};
\end{feynman}
\end{tikzpicture}}}}
\vcenter{\hbox{\scalebox{0.7}{\begin{tikzpicture}
\begin{feynman}
\vertex   (x);
\vertex [right=2.7cm of x]  (y);
\vertex [below=1.5cm of y]  (z);
\vertex [right=2cm of y]  (w);
\vertex [right=2cm of z]  (w1);
\vertex [left=2.7cm of z]  (x1);
\vertex [below=0.6cm of z] (invisible);
\vertex[below=0.75cm of y] (center);
\vertex [left=2.4cm of center] (center-left);
\vertex [above=0.3cm of center-left] (center-left-up);
\vertex [below=0.3cm of center-left] (center-left-down); 
\vertex [below=0.6cm of z] (invisible);
\diagram*{
 (x) --  [momentum={[arrow shorten=0.12mm,arrow distance=2mm]$p_1+\hbar w$}](y),
 (y) -- [momentum={[arrow shorten=0.12mm,arrow distance=2mm,xshift=0.3cm]$p_1+\hbar q$}](w),
 (x1) --  [momentum={[arrow shorten=0.12mm,arrow distance=2mm]$p_2-\hbar w$}](z),
 (z) --  [momentum={[arrow shorten=0.12mm,arrow distance=2mm,xshift=0.3cm]$p_2-\hbar q$}](w1),
 (z) --[graviton] (y),
 (z) --[color=white] (invisible),
};
\end{feynman}
\end{tikzpicture}}}}\right\}_{\mathcal{O}(\hbar^0)}
\end{align}
and an explicit evaluation indeed gives
\begin{equation}\label{super-cancel}
    \langle\Delta \mathbb{P}^\mu_1\rangle_{\text{\tiny 2PM,$A,\hbar^{-1}$}}+\langle\Delta \mathbb{P}^\mu_1\rangle_{\text{\tiny 2PM,$B,\hbar^{-1}$}}=0
\end{equation}

We now compare these Feynman diagrams from the KMOC-prescription with the 2PM evaluations using worldlines as in eq.~(\ref{dP-iterated-diagrams}) to find the following correspondence. First, for the contributions $\langle \mathbb{P}^\mu_1\rangle_{\text{\tiny 2PM,$Y$}}$ and $\langle \mathbb{P}^\mu_1\rangle_{\text{\tiny 2PM,$Y'$}}$ in eqs.~(\ref{dP-2PM-KMOC-Y}) and~(\ref{dP-2PM-KMOC-Y'})  we find\footnote{Note that these two worldline diagrams both have a symmetry factor of $1/2$.}
\begin{align}\label{dP-wl-star}
  \langle \mathbb{P}^\mu_1\rangle_{\text{\tiny 2PM,$Y$}}&=\frac{\pi G^2 m_2m_1^2}{4}\frac{\gamma^2+3}{\sqrt{\gamma^2-1}}\frac{\delta_i^\mu b_\perp^i}{(\vec{b}_\perp^2)^{3/2}}\\\nonumber&=G^2\int_{-\infty}^{\infty} d\tau \vcenter{\hbox{\scalebox{0.7}{\begin{tikzpicture}
\begin{feynman}
\node [circle,draw=black, fill=black,inner sep=1pt,text width=3mm] (x) {\textcolor{white}{$0$}};
\node [below=0.4cm of x ] (x1);
\vertex [above=0.75cm of x]  (center);
\node [above=0.75cm of center ]  (whaleross);
\node [left=0.5cm of whaleross,circle,draw=black, fill=white,inner sep=1pt,text width=3mm]  (y-left) {$0$};
\node [right=0.5cm of whaleross,circle,draw=black, fill=black,inner sep=1pt,text width=3mm]  (y-right){\textcolor{white}{$0$}};
\node [above=0.3cm of y-left ] (y1-left);
\node [left=0.3cm of y1-left ] (x2) {$\mu$};
\node [above=0.3cm of y-right ] (y1-right);
\diagram*{
(x) -- [graviton,thick,opacity=0.7] (center),
(x) -- [fermion] (center),
(center) -- [graviton,thick,opacity=0.7] (y-left),
(center) -- [fermion] (y-left),
(y-right) -- [graviton,thick,opacity=0.7] (center),
(y-right) -- [fermion] (center),
};
\end{feynman}
\end{tikzpicture}}}}
\end{align}
\begin{align}\label{dP-wl-star'}
   \langle \mathbb{P}^\mu_1\rangle_{\text{\tiny 2PM,$Y'$}}&=\frac{\pi G^2 m_2^2m_1}{4}\frac{\gamma^2+3}{\sqrt{\gamma^2-1}}\frac{\delta_i^\mu b_\perp^i}{(\vec{b}_\perp^2)^{3/2}}\\\nonumber &=G^2\int_{-\infty}^{\infty} d\tau \vcenter{\hbox{\scalebox{0.7}{\begin{tikzpicture}
\begin{feynman}
\node [circle,draw=black, fill=white,inner sep=1pt,text width=3mm] (x) {$0$};
\node [above=0.4cm of x ] (x1) {$\mu$};
\vertex [below=0.75cm of x]  (center);
\node [below=0.75cm of center ]  (whaleross);
\node [left=0.5cm of whaleross,circle,draw=black, fill=black,inner sep=1pt,text width=3mm]  (y-left){\textcolor{white}{$0$}};
\node [right=0.5cm of whaleross,circle,draw=black, fill=black,inner sep=1pt,text width=3mm]  (y-right){\textcolor{white}{$0$}};
\node [below=0.3cm of y-left ] (y1-left);
\node [below=0.3cm of y-right ] (y1-right);
\diagram*{
(center) -- [graviton,thick,opacity=0.7] (x),
(center) -- [fermion] (x),
(y-left) -- [graviton,thick,opacity=0.7] (center),
(y-left) -- [fermion] (center),
(y-right) -- [graviton,thick,opacity=0.7] (center),
(y-right) -- [fermion] (center),
};
\end{feynman}
\end{tikzpicture}}}}
\end{align}

These relations are diagrammatically very intuitive, and they match the correspondence with velocity cuts~\cite{Bjerrum-Bohr:2021din}. However, for the remaining contributions
the correspondence may appear less obvious. We discover that we need to sum the contributions to $\langle \mathbb{P}^\mu_1\rangle$ in eqs.~(\ref{dP-2PM-KMOC-Lam}) and~(\ref{dP-2PM-KMOC-B}) to get the first iterated worldline diagram in eq. (\ref{dP-iterated-diagrams}),
\begin{align}\label{dP-wl-it1}
    \langle \mathbb{P}^\mu_1\rangle_{\text{\tiny 2PM,$\Lambda$}}+\langle \mathbb{P}^\mu_1\rangle_{\text{\tiny 2PM,B.1}}&=-\frac{4\pi G^2 m_2^2m_1\gamma^2}{\sqrt{\gamma^2-1}}\frac{\delta^\mu_ib_\perp^i}{(\vec{b}_\perp^2)^{3/2}}+2\frac{G^2m_1m_2^2(2\gamma^2-1)^2}{\vec{b}_\perp^2(\gamma^2-1)^2}[\gamma u_2^\mu-u_1^\mu]\\\nonumber &=G^2\int_{-\infty}^{\infty} d\tau \vcenter{\hbox{\scalebox{0.7}{\begin{tikzpicture}
\begin{feynman}
\node [circle,draw=black, fill=white,inner sep=1pt,text width=3mm] (x) {$1$};
\node [above=0.4cm of x ] (x1) {$\mu$};
\node [below=1.5cm of x,circle,draw=black, fill=black,inner sep=1pt,text width=3mm]  (y){\textcolor{white}{$0$}};
\node [below=0.3cm of y ] (y1);
\diagram*{
 (y) -- [graviton,thick,opacity=0.7] (x),
 (y) -- [fermion] (x),
};
\end{feynman}
\end{tikzpicture}}}}
\end{align}
and likewise, the sum of eqs. (\ref{dP-2PM-KMOC-V}) and (\ref{dP-2PM-KMOC-B2}) gives the other first iterated worldline diagram in eq. (\ref{dP-iterated-diagrams})
\begin{align}\label{dP-wl-it2}
    \langle \mathbb{P}^\mu_1\rangle_{\text{\tiny 2PM,V}}+\langle \mathbb{P}^\mu_1\rangle_{\text{\tiny 2PM,B.2}}&=-\frac{4\pi G^2 m_2m_1^2\gamma^2}{\sqrt{\gamma^2-1}}\frac{\delta^\mu_ib_\perp^i}{(\vec{b}_\perp^2)^{3/2}}+2\frac{G^2m_1^2m_2(2\gamma^2-1)^2}{\vec{b}_\perp^2(\gamma^2-1)^2}[u_2^\mu-\gamma u_1^\mu]\\\nonumber &=G^2\int_{-\infty}^{\infty} d\tau \vcenter{\hbox{\scalebox{0.7}{\begin{tikzpicture}
\begin{feynman}
\node [circle,draw=black, fill=white,inner sep=1pt,text width=3mm] (x) {$0$};
\node [above=0.4cm of x ] (x1) {$\mu$};
\node [below=1.5cm of x,circle,draw=black, fill=black,inner sep=1pt,text width=3mm]  (y) {\textcolor{white}{$1$}};
\node [below=0.3cm of y ] (y1);
\diagram*{
 (y) -- [graviton,thick,opacity=0.7] (x),
 (y) -- [fermion] (x),
};
\end{feynman}
\end{tikzpicture}}}}
\end{align}
Looking at the diagrams, we observe that the worldline iterations can be viewed as producing, to this order, two contributions:
one is the analog of higher order scalar-gravition vertices in Feynman diagrams, the other is the classical contributions from derivatives acting on the $\delta$-function constraints
(and additional power of $q^2$). We are not used to thinking of such terms as being iterations in a conventional Feynman diagram analysis but it hints at the possibility of a re-organization (or merging) of the two formalisms where one chooses the most convenient representation to enumerate contributions. It should also be noted that the split in the box and crossed box diagrams is easily
identified by which $\delta$-function constraint the derivative acts upon in the KMOC formulation. 
This is what gives rise to the asymmetric powers of $m_1m_2^2$ and $m_2m_1^2$, respectively.
We have not explored the diagrammatic correspondence in the alternative
Worldline Quantum Field Theory~\cite{Mogull:2020sak,Jakobsen:2022psy} but one might expect that the correspondence could be even simpler in that framework.
 Indeed, such simplifications have been noticed~\cite{Comberiati:2022ldk} in the context of off-shell currents.

\section{Conclusion}\label{sec:conclusion}

The KMOC and worldline formulations of classical general relativity are both used to compute observables based on the {\em in-in} formalism
of quantum field theory. It is thus natural to ask: what is the relationship between the two? This issue becomes particularly relevant when one
considers dissipation in terms of gravitational radiation. The worldline formalism leads naturally to retarded (or advanced) propagators
in this context, while the KMOC formalism evaluates matrix elements with the standard $i\epsilon$-prescription of Feynman propagators.
In this paper we have shown that there is no contradiction, and we have provided a formal derivation of the worldline formulation from the 
classical limit of the KMOC formalism.
One of the features of the derivation is that the initial conditions that need to be imposed on the trajectories in the worldline formulation, can be viewed as originating directly from the initial-state wave-functions that specify the initial conditions of the system in KMOC.\\

One of the advantages of the worldline formulations is that they can work with the $\hbar \to 0$ limit from the outset. The formal equivalence to the classical
limit of the KMOC formalism demonstrates implicitly the exact cancellation of all superclassical terms if the KMOC matrix elements are evaluated by
means of the standard Born expansion of the $S$-matrix. The interesting question of what happens to the equivalence at the quantum level remains 
unanswered by the present analysis where we repeatedly discarded quantum corrections. Both formalisms should remain valid (and hence equivalent)
also when including quantum corrections. Although it is far from obvious how, all discarded quantum terms must thus conspire to keep the equivalence to all orders
in $\hbar$. 

Once this equivalence has been established, it is of interest to see in detail how computations in the two different approaches compare.
We have illustrated this by comparing both tree-level and one-loop results for the momentum kick of two massive scalars scattering off
each other. Interestingly, and this is apparent already at tree level, the contributions come from entirely different directions, eventually leading
to identical results. At one-loop level one can clearly identify which parts of the calculations are in one-to-one correspondence with
each other. The link between the two is most transparent if one organizes the quantum field theory calculation in terms of velocity cuts
on the massive lines. It would be interesting to generalize this to any loop order and see if simplifications can be introduced in the worldline
calculations based on what we know from the classical limit of amplitude calculations in the KMOC setting.

\subsection*{Acknowledgements}
We thank Jan Plefka for discussions. P.H.D. and P.V. would like to thank CERN Theory Department and LAPTh, respectively, for the hospitality during the
completion of this work. P.V. is grateful to the Mainz
Institute for Theoretical Physics (MITP) of the Cluster of Excellence
PRIMA$^+$ (Project ID 39083149), for its hospitality and its partial
support during the completion of this work. 
The work of P.H.D. was supported in part by DFF grant 0135-00089A,
the work of E.R.H. was supported by the Rozenthal Foundation and
ERC Starting Grant No. 757978 from the European Research Council, and 
the research of P.V. has received funding from the ANR grant ``SMAGP''
ANR-20-CE40-0026-01. 

\appendix

\section{Worldline Feynman rules}\label{WEFT-Feynman-rules}
The derivation of the Feynman rules for the variation of the effective action in eq.~(\ref{dP_1-Seff}) in the {\em in-in} formalism has been treated previously in, {\em e.g.},~\cite{Kalin:2022hph,Galley:2009px}. We will here briefly repeat the main steps and define the notation we are using this paper.

The explicit equation for the interacting part of the effective action is given by
\begin{align}
    e^{\frac{i}{\hbar}S_\text{eff}^{(\text{int})}[z_j^{(1)},z_j^{(2)}]}=\int\limits_{\text{in-in}}Dh^{(i)} e^{\frac{i}{\hbar}(S_h[h^{(1)},J^{(1)}]+S_{p}^{(\text{int})}[z^{(1)};h^{(1)}]-S_h[h^{(2)},J^{(2)}]-S_{p}^{(\text{int})}[z^{(2)};h^{(2)}])}
\end{align}
where we have introduced the following notation for the interacting part of the point-particle action,
\begin{align}
    S_{p}^{(\text{int})}[z;h]=-\sqrt{8\pi G}\sum_j m_j\int{d\tau_j }h_{\alpha\beta}(z_j)\dot{z}_j^\alpha\dot{z}_j^\beta\\\nonumber
\end{align}

\noindent Next, we write the purely gravitational part of the action as a sum of the kinetic and interacting parts, and we include a source $J_{\mu\nu}$ for the gravitational field,
\begin{equation}
    S_h[h,J]=-\frac{1}{2} M^{\alpha\beta\mu\nu}\int d^4x h_{\alpha\beta}\Box h_{\mu\nu}+S_h^{(\text{int})}[h]+\int d^4x J^{\mu\nu}h_{\mu\nu}
\end{equation}
\noindent where $M^{\rho\lambda,\alpha\beta}=\eta^{\rho(\alpha}\eta^{\beta)\lambda}-\frac{1}{2}\eta^{\rho\lambda}\eta^{\beta\alpha}$. Then we do a variable change to the Keldysh-variables for the gravitational field~\cite{Galley:2009px}. They are related to the $\{1,2\}$-variables used above by
\begin{equation}
    \left[ \begin{array}{c}
h^{(1)} \\ 
h^{(2)} \end{array}
\right]=\left[ \begin{array}{cc}
\frac{1}{2}~~ & 1 \\ 
-\frac{1}{2}~~ & 1 \end{array}
\right]\left[ \begin{array}{c}
h^{\left(-\right)} \\ 
h^{\left(+\right)} \end{array}
\right]
\end{equation}
We define the interacting part of the action in the Keldysh-basis for the gravitational field as (we have kept the worldline variables in the non-Keldysh basis, since we are not integrating over them)
\begin{align}\label{Gamma}
    \mathcal{S}^{(\text{int})}[z^{(1)},z^{(2)},h^{(\pm)}]\equiv &S_{p+h}^{(\text{int})}[z^{(1)};h^{(+)}+\frac{1}{2}h^{(-)}]-S_{p+h}^{(\text{int})}[z^{(2)};h^{(+)}-\frac{1}{2}h^{(-)}],
\end{align}
\noindent where $S_{p+h}^{(\text{int})}[z;h]=S_{p}^{(\text{int})}[z;h]+S_{h}^{(\text{int})}[h]$. Also switching to the Keldysh-basis for the sources $J_{\mu\nu}^{(\pm)}$, the effective action can be evaluated as follows
\begin{align}\label{Seff-W}
    e^{\frac{i}{\hbar}S_\text{eff}^{(\text{int})}[z_j^{(1)},z_j^{(2)}]}=&\mathcal{N}\left.\text{exp}\left\{\frac{i}{\hbar}\mathcal{S}^{(\text{int})}\left[z_j^{(1)},z_j^{(2)},-i\hbar\frac{\delta}{\delta J^{(\mp)}}\right]\right\}e^{\frac{i}{\hbar}W[J^{(\pm)}]}\right|_{J_{\mu\nu}^{(\pm)}=0}
\end{align}
where
\begin{equation}\label{W}
    W[J^{(\pm)}]\equiv \frac{M^{\rho\lambda,\alpha\beta}}{2}\int \frac{d^4k}{(2\pi)^4}\left[ \begin{array}{c}
J^{(-)}_{\rho\lambda}(-k) \\ 
J^{(+)}_{\rho\lambda}(-k)  \end{array}
\right]^T\left[ \begin{array}{cc}
i\pi \delta (k^2) & \frac{1}{(k^0+i\epsilon)^2-\vec{k}^2} \\ 
\frac{1}{(k^0-i\epsilon)^2-\vec{k}^2}  & 0 \end{array}
\right]\left[ \begin{array}{c}
J^{(-)}_{\alpha\beta}(k) \\ 
J^{(+)}_{\alpha\beta}(k)  \end{array}
\right]\
\end{equation}
The non-zero diagonal element can be shown
not to contribute to classical  observables~\cite{Jakobsen:2022psy,Kalin:2022hph} since it cannot occur in diagrams without loops or, alternatively, multiple sinks. So effectively we only need to worry about the retarded propagator
\begin{equation}
    W[J^{(\pm)}]=M^{\rho\lambda,\alpha\beta}\int \frac{d^4k}{(2\pi)^4}\frac{J^{(-)}_{\rho\lambda}(-k)J^{(+)}_{\alpha\beta}(k)}{(k^0+i\epsilon)^2-\vec{k}^2} 
    \end{equation}
after we have discarded all quantum effects in $W[J^{(\pm)}]$.

The explicit expression for the momentum change can now be obtained by inserting the effective action calculated according to~(\ref{Seff-W}) into equation~(\ref{dP_1-Seff}). The diagrammatic rules for evaluating the momentum change can be worked out from~(\ref{Seff-W}). Apart from gravity-self-interactions, we get from~(\ref{Gamma}) two types of interaction terms. After Fourier transforming the graviton field (ignoring convergence subtleties), they read
\begin{align}
    -m_j\sqrt{8\pi G}\int d\tau \int \frac{d^4k}{(2\pi)^4} &h^{(+)}_{\alpha\beta}(k)(e^{-ik\cdot z_j^{(1)}}\dot{z}^{\alpha{(1)}}_j\dot{z}^{\beta{(1)}}_j-e^{-ik\cdot z_j^{(2)}}\dot{z}^{\alpha{(2)}}_j\dot{z}^{\beta{(2)}}_j)\\
    -\frac{1}{2}m_j\sqrt{8\pi G}\int d\tau \int \frac{d^4k}{(2\pi)^4} &h^{(-)}_{\alpha\beta}(k)(e^{-ik\cdot z_j^{(1)}}\dot{z}^{\alpha{(1)}}_j\dot{z}^{\beta{(1)}}_j+e^{-ik\cdot z_j^{(2)}}\dot{z}^{\alpha{(2)}}_j\dot{z}^{\beta{(2)}}_j)
\end{align}

It is customary to think of the latter of these terms as a 'source' and the former as a 'sink' for the following reason: we only care about classical diagrams that contribute after we take the variation of the effective action with respect to $z^{\alpha{(1)}}_j$ and subsequently set the two sets of variables equal $z^{\alpha{(1)}}_j=z^{\alpha{(2)}}_j=r^{\alpha}_j$. Only diagrams with exactly one sink-vertex will survive this operation, and only when we take the variation of only that sink-vertex. Thus, when calculating diagrams contributing to~(\ref{dP_1-Seff}), we may operate with 
$$
-(-i\hbar)\eta^{\mu\nu} \left.\frac{\delta}{\delta z^{\nu(1)}_j(\tau)}\right|_{z^{\nu(i)}_j=r^{\nu}_j}
$$
on the sink-term in advance and set $z^{\nu(i)}=r^{\nu}$ on the source term to obtain the vertex factors given below. Here, the factor of $-i\hbar$ comes from the left-hand side of (\ref{Seff-W}). As usual we define the diagrams such that sinks/sources corresponding to particle 1 are written on top of vertices corresponding to particle 2

\begin{align}\label{dP_diagrams3}
&\vcenter{\hbox{\scalebox{0.9}{\begin{tikzpicture}
\begin{feynman}
\node (x);
\node [below=1cm of x,circle,draw=white, fill=black,inner sep=1pt,text width=3mm]  (y);
\node [above=0.2cm of x ] (x1) {$\alpha\beta$};
\node [below=0.3cm of y ] (y1);
\node [right=3.7cm of y] (y2) {$\equiv -\frac{i\sqrt{8\pi}m_2}{\hbar}\int_{-\infty}^{\infty} d\tau' e^{-i k\cdot r_2(\tau')} \dot{r}_2^{\alpha}(\tau')\dot{r}_2^{\beta}(\tau')$};
\diagram*{
 (y) -- [graviton,thick,opacity=0.7] (x),
 (y) -- [fermion, rmomentum={$\hbar k$}] (x),
};
\end{feynman}
\end{tikzpicture}}}}\\\nonumber
&\vcenter{\hbox{\scalebox{0.9}{\begin{tikzpicture}
\begin{feynman}
\node (x);
\node [below=1cm of x,circle,draw=black, fill=white,inner sep=1pt,text width=3mm]  (y);
\node [above=0.2cm of x ] (x1) {$\alpha\beta$};
\node [below=0.4cm of y ] (y1) {$\mu$};
\node [right=6cm of y] (y2) {$\equiv \sqrt{8\pi}m_2\left[ ik^\mu e^{-i k\cdot r_2(\tau)} \dot{r}_2^{\alpha}(\tau)\dot{r}_2^{\beta}(\tau)+2\frac{d}{d\tau}[e^{-i k\cdot r_2(\tau)} \eta^{\mu(\alpha}\dot{r}_2^{\beta)}(\tau)]\right]$};
\diagram*{
 (y) -- [graviton,thick,opacity=0.7] (x),
 (y) -- [anti fermion, rmomentum={$\hbar k$}] (x),
};
\end{feynman}
\end{tikzpicture}}}}
\end{align}
\begin{align}\label{dP_diagrams2}
&\vcenter{\hbox{\scalebox{0.9}{\begin{tikzpicture}
\begin{feynman}
\node (x);
\node [above=1cm of x,circle,draw=white, fill=black,inner sep=1pt,text width=3mm]  (y);
\node [below=0.2cm of x ] (x1) {$\alpha\beta$};
\node [above=0.3cm of y ] (y1);
\node [right=3.7cm of y] (y2) {$\equiv -\frac{i\sqrt{8\pi}m_1}{\hbar}\int_{-\infty}^{\infty} d\tau' e^{-i k\cdot r_1(\tau')} \dot{r}_1^{\alpha}(\tau')\dot{r}_1^{\beta}(\tau')$};
\diagram*{
 (y) -- [graviton,thick,opacity=0.7] (x),
 (y) -- [fermion, rmomentum={$\hbar k$}] (x),
};
\end{feynman}
\end{tikzpicture}}}}\\\nonumber
&\vcenter{\hbox{\scalebox{0.9}{\begin{tikzpicture}
\begin{feynman}
\node (x);
\node [above=1cm of x,circle,draw=black, fill=white,inner sep=1pt,text width=3mm]  (y);
\node [below=0.2cm of x ] (x1) {$\alpha\beta$};
\node [above=0.4cm of y ] (y1) {$\mu$};
\node [right=6cm of y] (y2) {$\equiv \sqrt{8\pi}m_1\left[ ik^\mu e^{-i k\cdot r_1(\tau)} \dot{r}_1^{\alpha}(\tau)\dot{r}_1^{\beta}(\tau)+2\frac{d}{d\tau}[e^{-i k\cdot r_1(\tau)} \eta^{\mu(\alpha}\dot{r}_1^{\beta)}(\tau)]\right]$};
\diagram*{
 (y) -- [graviton,thick,opacity=0.7] (x),
 (y) -- [anti fermion, rmomentum={$\hbar k$}] (x),
};
\end{feynman}
\end{tikzpicture}}}}
\end{align}
and the propagator is given by
\begin{align}\label{dP_diagramsX}
&\vcenter{\hbox{\scalebox{0.9}{\begin{tikzpicture}
\begin{feynman}
\node (x);
\node [right=1cm of x]  (y);
\node [left=0.2cm of x ] (x1) {$\alpha\beta$};
\node [right=0.2cm of y ] (y1) {$\mu\nu$};
\node [right=1.5cm of y1] (y2) {$\equiv \frac{i\hbar M^{\alpha\beta\mu\nu}}{(k^0+i\epsilon)^2-\vec{k}^2}$};
\diagram*{
 (y) -- [graviton,thick,opacity=0.7] (x),
 (y) -- [anti fermion, rmomentum={$\hbar k$}] (x),
};
\end{feynman}
\end{tikzpicture}}}}
\end{align}
Momentum is conserved at each vertex and in the end all momentum variables are integrated over. Notice that we have factored out an appropriate power of the coupling constant $G$; however, the vertex factors contain additional factors of $G$, through their dependence on the classical trajectories as illustrated in the main text.

\section{Classical limit of the worldline}\label{classical-worldline-representation}
In the derivation presented above we used the worldline representation for the solution to
\beq
    -\left(\frac{1}{\sqrt{-g}}{\partial }_{\mu }(\sqrt{-g}g^{\mu \nu }{\partial }_{\nu })+\frac{m^2_j}{{\hbar }^2}-i\epsilon\right)\mathrm{\Delta }_j\left(x,y;h\right)=\frac{{\delta }^4\left(x-y\right)}{\sqrt{-g}} ~.
\eeq
We stated in the main text that the result may be written as follows~\cite{Bastianelli:2002fv,Bastianelli:1998jm,Mogull:2020sak}\footnote{Up to regularization-dependent non-covariant counter-terms, see ref.~\cite{Bastianelli:1998jm} for a discussion of this issue.}

\begin{equation}\label{Schwinger-feynman-B}
    \mathrm{\Delta }_j\left(x,y;h\right)=\mathcal{N}\int^{\infty }_{0}{d T }\int^{z\left( T \right)=y}_{z\left(0\right)=x}{\mathcal{D}z}e^{-\frac{i}{\hbar }\frac{m_j}{2}\int^{ T }_{0}{ds}\left\{g_{\mu \nu }\left(z\right){\dot{z}}^{\mu }{\dot{z}}^{\nu }-\frac{{\hbar }^2}{4m_j^2}R(z)+1\right\}}
\end{equation}
where 
\beq
\mathcal{D}z\equiv \prod_{0<s<T}{d^4z(s)\sqrt{-g(z(s))}} ~.
\eeq 
We wish to simplify this as much as possible by discarding effects that will only become
of importance at the quantum level. Analogous considerations
have already been described in ref.~\cite{Mogull:2020sak} but we provide a slightly different approach.

Clearly, the $-\frac{{\hbar }^2}{4m^2}R(z)$-term 
in the exponent will not contribute in the classical limit and we hence omit it. Similarly, the factor of
$\sqrt{-g(z(s))}$ in the measure exponentiates to a term that is manifestly of quantum origin on account of the
additional factor of $\hbar$ and we can therefore ignore it if we focus on the classical limit.  Next, we
make the change of variables $s\to u= s/ T$. This allows us to simplify~(\ref{Schwinger-feynman-B}) to
\begin{equation}\label{Schwinger-feynmann-u}
    \mathrm{\Delta }_j\left(x,y;h\right)=\mathcal{N}\int^{z\left(1\right)=y}_{z\left(0\right)=x}{\mathcal{D}z}\int^{\infty }_0{dT}e^{\frac{i}{\hbar }S_p\left[z;h;T\right]}
\end{equation}
where 
\begin{equation}\label{Sp}
    S_p[z;h;T]\equiv -\frac{m_j}{2}\int^1_0{du}\left\{\frac{1}{T}g_{\mu \nu }\left(z\right){\frac{dz}{du}}^{\mu }{\frac{dz}{du}}^{\nu }+T\right\}~.
\end{equation}
In the classical limit we can evaluate the $T$-integral by the principle of stationary phase,
\begin{equation}\label{EOM-T}
    \left.\frac{\partial}{\partial T}S_p[z;h;T]\right|_{T=T_\text{cl}[z,h]}=0
\end{equation}
which leads to the condition
\begin{equation}\label{T0-def}
    T_\text{cl}[z,h]=\left\{\int^1_0{du}g_{\mu \nu }(z){\frac{dz}{du}}^{\mu }{\frac{dz}{du}}^{\nu}\right\}^{1/2}
\end{equation}
Because the saddle-point condition on $T$ implies that the equations of motion for $z$ and $h$ are independent of the
functional dependence of $T_\text{cl}[z,h]$ on $z$ and $h$, we can treat $T_\text{cl}$ as evaluated at the solutions to classical equations
of motion for $z$ and $h$, which we denote with subscript $\small \text{cl}$ for the present purposes. When analysing the equations of motion for $z$, one finds that 
\beq
\frac{d}{du}\left(g_{\mu \nu }(z_\text{cl}(u))\frac{dz_\text{cl}^{\mu}}{du}\frac{dz_\text{cl}^{\nu}}{du}\right)=0 ~,
\eeq
so $T_\text{cl}[z_\text{cl},h]$ is (the square root of) an integral of a constant. The integral is thus trivial. From~(\ref{T0-def}) we can then identify $T_\text{cl}[z_\text{cl},h_\text{cl}]$ as
\begin{equation}\label{T0-cl}
    T_\text{cl}\equiv T_\text{cl}[z_\text{cl},h_\text{cl}]=\left\{g_{\mu \nu}(z_\text{cl})\frac{dz_\text{cl}^{\mu}}{du}\frac{dz_\text{cl}^{\nu}}{du}\right\}^{1/2}
\end{equation}
We now change variables back to from 
$u$ to $\tau =uT_\text{cl}$ and write the action as
\begin{align}\label{Sp-T0}
    S_p\left[z;h\right]\equiv S_p\left[z;h;T_\text{cl}\right]&=-\frac{m_j}{2}\int^{T_\text{cl}}_0{d\tau }\left\{g_{\mu\nu}(z)\frac{dz^{\mu }}{d\tau }\frac{dz^{\nu }}{d\tau }+1\right\}
\end{align}
which gives us the classical part of the Green function
\beq\label{Delta-classical-final}
\Delta_j(x,y;h)= \mathcal{N}\int^{z ({\tau }_{\text{f}})= y}_{z ({\tau }_{\text{in}})= x}{\mathcal{D}z}e^{\frac{i}{\hbar}\int^{{\tau }_{\text{f}}}_{{\tau }_{\text{in}}}{d\tau }L_j[z,\dot{z},h]}
\eeq
where $\tau_{f}-\tau_\text{in}=T_\text{cl}$ is still the classical proper time as defined in~(\ref{T0-cl}). We have absorbed the normalization factor from the saddle-point integration
over $T$ into the overall constant $\mathcal{N}$ and
\begin{equation}
    L_j\left[z,\dot{z},h\right]\equiv-\frac{m_j}{2}\left[g_{\mu \nu }\left(z\right)\frac{dz^{\mu }}{d\tau }\frac{dz^{\nu }}{d\tau }+1\right],
\end{equation}
is the Polyakov form of the classical Lagrangian for a scalar particle on curved spacetime.\\

\noindent As a side-note,~(\ref{T0-cl}) also implies that the classical path satisfies the useful relation 
\begin{equation}\label{v^2=1}
    1=g_{\mu \nu }(z_\text{cl})\frac{dz_\text{cl}^{\mu}}{d\tau}\frac{dz_\text{cl}^{\nu}}{d\tau},
\end{equation}
\noindent  which implies that the action evaluated along the classical path is proportional to the proper-time difference
\begin{equation}
     S_p\left[z_\text{cl};h\right]=-m_j\int_{\tau_\text{in}}^{\tau_\text{f}} d\tau
\end{equation}
in agreement with what we would expect from classical mechanics.

\section{Derivatives with respect to worldline end-points}
\label{derivative-worldline}

In this appendix we go through the technical details on how to take derivatives with respect to the end-points of the worldline action
in the classical limit.
We begin by considering one derivative with respect to the upper end-point of~(\ref{Schwinger-feynman-classical})
\begin{equation}\label{eq:2.71}
    \left(-i\hbar {\partial }_{{y}^\mu}\right)\Delta(x,y;h)
\end{equation}
Here we have dropped the particle label. We define $z_\text{cl}[x,y;\tau]$ to be solution to the classical equation of motion, {\em i.e.}, it satisfies
\begin{equation}\label{eq:2.72}
    {\left[\frac{\partial L\left[z,\dot{z},h\right]}{\partial z^{\mu }}-\frac{d}{d\tau }\frac{\partial L\left[z,\dot{z},h\right]}{\partial {\dot{z}}^{\mu }}\right]}_{z=z_\text{cl}}=0,
\end{equation}
subject to the boundary condition $z^{\mu }_\text{cl}[x,y;{\tau }_{\mathrm{in}}]=x$ and $z^{\mu }_\text{cl}[x,y;{\tau }_{\mathrm{f}}]=y$. We next shift variables in the path integral~(\ref{Schwinger-feynman-classical}) as follows
\begin{equation}\label{eq:2.73}
    z^{\mu }(\tau )=z^{\mu }_\text{cl}[x,y;\tau]+{\zeta }^{\mu }(\tau )
\end{equation}
It follows that the boundary conditions on the new variable are ${\zeta }^{\mu }(\tau_{\text{f}} )={\zeta }^{\mu }(\tau_{\text{in}} )=0$. Now we can write the derivative as
\begin{eqnarray}\label{eq:2.74}
   && (-i\hbar {\partial }_{y^{\mu }})\Delta(x,y;h)=\int^{\zeta \left({\tau }_{\mathrm{f}}\right)=0}_{\zeta \left({\tau }_{\mathrm{in}}\right)=0}{\mathcal{D}\zeta }(-i\hbar {\partial }_{y^{\mu }})e^{\frac{i}{\hbar }\int^{{\tau }_{\mathrm{f}}}_{{\tau }_{\mathrm{in}}}{d\tau }L\left[z_\text{cl}[x,y;\tau]+\zeta ,{\dot{z}}_\text{cl}[x,y;\tau]+\dot{\zeta },h\right]} \cr
    &&=\int^{\zeta \left({\tau }_{\mathrm{f}}\right)=0}_{\zeta \left({\tau }_{\mathrm{in}}\right)=0}{\mathcal{D}\zeta }\int^{{\tau }_{\mathrm{f}}}_{{\tau }_{\mathrm{in}}}{d\tau }\left[\frac{\partial L\left[z,\dot{z},h\right]}{\partial z^{\nu }}\ {\partial }_{y^{\mu }}z^{\nu }_\text{cl}+\frac{\partial L\left[z,\dot{z},h\right]}{\partial {\dot{z}}^{\nu }}{\partial }_{y^{\mu }}{\dot{z}}^{\nu }_\text{cl}\right]e^{\frac{i}{\hbar }\int^{{\tau }_{\mathrm{f}}}_{{\tau }_{\mathrm{in}}}{d\tau }L} \cr
    &&=\int^{\zeta \left({\tau }_{\mathrm{f}}\right)=0}_{\zeta \left({\tau }_{\mathrm{in}}\right)=0}{\mathcal{D}\zeta }\left\{{\left[\frac{\partial L}{\partial {\dot{z}}^{\nu }}{\partial }_{y^{\mu }}z^{\nu }_\text{cl}\right]}^{\tau ={\tau }_{\mathrm{f}}}_{\tau ={\tau }_{\mathrm{in}}}+\int^{{\tau }_{\mathrm{f}}}_{{\tau }_{\mathrm{in}}}{d\tau }{\partial }_{y^{\mu }}z^{\nu }_\text{cl}{\left[\frac{\partial L}{\partial z^{\nu }}\ -\frac{d}{d\tau }\frac{\partial L}{\partial {\dot{z}}^{\nu }}\right]}_{z=z_\text{cl}+\zeta }\right\}e^{\frac{i}{\hbar }\int^{{\tau }_{\mathrm{f}}}_{{\tau }_{\mathrm{in}}}{d\tau }L} \cr
    &&=\int^{\zeta \left({\tau }_{\mathrm{f}}\right)=0}_{\zeta \left({\tau }_{\mathrm{in}}\right)=0}{\mathcal{D}\zeta }\left\{{\left[\frac{\partial L}{\partial {\dot{z}}^{\nu }}{\partial }_{y^{\mu }}z^{\nu }_\text{cl}\right]}^{\tau ={\tau }_{\mathrm{f}}}_{\tau ={\tau }_{\mathrm{in}}}+\mathcal{O}(\hbar )\right\}e^{\frac{i}{\hbar }\int^{{\tau }_{\mathrm{f}}}_{{\tau }_{\mathrm{in}}}{d\tau }L}
\end{eqnarray}
We have here used the fact that the last term in the third line is manifestly proportional to $\hbar$ (and higher orders). This follows from an exact Schwinger-Dyson equation since the integrand is proportional to the equations of motion.

Using ${\partial }_{y^{\mu }}z^{\nu }_\text{cl}[x,y;{\tau }_{\mathrm{in}}]={\partial }_{y^{\mu }}x^{\nu }=0$ and ${\partial }_{y^{\mu }}z^{\nu }_\text{cl}[x,y;{\tau }_{\mathrm{f}}]={\partial }_{y^{\mu }}y^{\nu }={\delta }^{\nu }_{\mu }$ we get from the last line in equation~(\ref{eq:2.74})
\begin{equation}\label{eq:2.75}
    \left(-i\hbar {\partial }_{y^{\mu }}\right)\Delta(x,y;h)=\int^{\zeta \left({\tau }_{\mathrm{f}}\right)=0}_{\zeta \left({\tau }_{\mathrm{in}}\right)=0}{\mathcal{D}\zeta }{\left.\frac{\partial L}{\partial {\dot{z}}^{\mu }}\right|}_{\tau ={\tau }_{\mathrm{f}}}e^{\frac{i}{\hbar }\int^{{\tau }_{\mathrm{f}}}_{{\tau }_{\mathrm{in}}}{d\tau }L}
\end{equation}
Had we instead differentiated with respect to $x$, only the lower boundary would survive, ${\partial }_{x^{\mu }}z^{\nu }_\text{cl}[x,y;{\tau }_{\mathrm{in}}]={\partial }_{x^{\mu }}x^{\nu }={\delta }^{\nu }_{\mu }$, so we would have
\begin{equation}\label{eq:2.76}
    \left(-i\hbar {\partial }_{x^{\mu }}\right)\Delta(x,y;h)=\int^{\zeta \left({\tau }_{\mathrm{f}}\right)=0}_{\zeta \left({\tau }_{\mathrm{in}}\right)=0}{\mathcal{D}\zeta }\left(-{\left.\frac{\partial L}{\partial {\dot{z}}^{\mu }}\right|}_{\tau ={\tau }_{\mathrm{in}}}\right)e^{\frac{i}{\hbar }\int^{{\tau }_{\mathrm{f}}}_{{\tau }_{\mathrm{in}}}{d\tau }L}
\end{equation}

Now let us explore what happens when we take one more derivative of~(\ref{eq:2.75})
\begin{eqnarray}\label{eq:2.77}
   && (-i\hbar {\partial }_{y^{\mu }})(-i\hbar {\partial }_{y^{\nu }})\Delta(x,y;h) \cr && =\int^{\zeta \left({\tau }_{\mathrm{f}}\right)=0}_{\zeta \left({\tau }_{\mathrm{in}}\right)=0}{\mathcal{D}\zeta }\left\{{\left.\frac{\partial L}{\partial {\dot{z}}^{\mu }}\right|}_{\tau ={\tau }_{\mathrm{f}}}{\left.\frac{\partial L}{\partial {\dot{z}}^{\nu }}\right|}_{\tau ={\tau }_{\mathrm{f}}}-i\hbar {\partial }_{y^{\nu }}{\left.\frac{\partial L}{\partial {\dot{z}}^{\mu }}\right|}_{\tau ={\tau }_{\mathrm{f}}}\right\}e^{\frac{i}{\hbar }\int^{{\tau }_{\mathrm{f}}}_{{\tau }_{\mathrm{in}}}{d\tau }L}
\end{eqnarray}
The second term is seen to be subleading in $\hbar $. In general, when we take any number of derivatives
$\left(-i\hbar {\partial }_{y^{\mu }}\right)$ is replaced by ${\left.\frac{\partial L}{\partial {\dot{z}}^{\mu }}\right|}_{\tau ={\tau }_{\mathrm{f}}}$ to leading order in $\hbar $ and $\left(-i\hbar {\partial }_{x^{\mu }}\right)$ is replaced by $-{\left.\frac{\partial L}{\partial {\dot{z}}^{\mu }}\right|}_{\tau ={\tau }_{\mathrm{in}}}$ to leading order in $\hbar $. Undoing the change of variables in~(\ref{eq:2.73}), we can write the result as
\begin{equation}\label{eq:2.78}
    {\left(-i\hbar {\partial }_y\right)}^n\Delta(x,y;h)=\int^{z\left({\tau }_{\mathrm{f}}\right)=y}_{z\left({\tau }_{\mathrm{in}}\right)=x}{\mathcal{D}z}\left\{{\left({\left.\frac{\partial L}{\partial {\dot{z}}}\right|}_{\tau ={\tau }_{\mathrm{f}}}\right)}^n+\mathcal{O}(\hbar )\right\}e^{\frac{i}{\hbar }\int^{{\tau }_{\mathrm{f}}}_{{\tau }_{\mathrm{in}}}{d\tau }L}
\end{equation}
and by exactly the same arguments
 \begin{equation}\label{eq:2.79}
  {\left(-i\hbar {\partial }_x\right)}^n\Delta(x,y;h)=\int^{z\left({\tau }_{\mathrm{f}}\right)=y}_{z\left({\tau }_{\mathrm{in}}\right)=x}{\mathcal{D}z}\left\{{\left(-{\left.\frac{\partial L}{\partial {\dot{z}}}\right|}_{\tau ={\tau }_{\mathrm{in}}}\right)}^n+\mathcal{O}(\hbar )\right\}e^{\frac{i}{\hbar }\int^{{\tau }_{\mathrm{f}}}_{{\tau }_{\mathrm{in}}}{d\tau }L}
 \end{equation}
which were the results quoted in the main text.

\end{document}